\documentclass[lettersize,journal]{IEEEtran}
\usepackage{amsmath,amsfonts}
\usepackage{algorithmic}
\usepackage{algorithm}
\usepackage{array}
\usepackage[caption=false,font=normalsize,labelfont=sf,textfont=sf]{subfig}
\usepackage{textcomp}
\usepackage{stfloats}
\usepackage{url}
\usepackage{verbatim}
\usepackage{graphicx}
\usepackage{booktabs}%
\usepackage{multirow}%
\usepackage{tabularx}
\usepackage[
  style=ieee,
  backend=biber,
  abbreviate=true,
  isbn=false,
  doi=false,
  url=false,
  eprint=false,
]{biblatex}

\usepackage{balance}

\usepackage{etoolbox}
\newcommand{\tablefontsize}{\footnotesize}
\AtBeginEnvironment{table}{\tablefontsize}
\AtBeginEnvironment{longtable}{\tablefontsize}

\renewbibmacro*{journal}{%
  \iffieldundef{shortjournal}
    {\printfield{journaltitle}}
    {\printfield{shortjournal}}%
}

\renewbibmacro*{journal+issuetitle}{%
  \usebibmacro{journal}%
  \setunit*{\addcomma\space}%
  \iffieldundef{series}
    {}
    {\newunit
     \printfield{series}%
     \setunit{\addspace}}%
  \usebibmacro{volume+number+eid}%
  \setunit{\addspace}%
  \usebibmacro{issue+date}%
  \setunit{\addcolon\space}%
  \usebibmacro{issue}%
  \newunit}

\DeclareFieldFormat{pages}{#1}

\renewbibmacro*{volume+number+eid}{%
  \printfield{volume}%
  \setunit*{\addnbspace}%
  \printfield{number}%
  \setunit{\addcomma\space}%
  \printfield{eid}%
}

\DeclareFieldFormat[online]{title}{\mkbibquote{#1}}

\renewbibmacro*{volume+number+eid}{%
  \printfield{volume}%
  \setunit*{\addnbspace}%
  \printfield{number}%
  \setunit{\addcomma\space}%
  \printfield{eid}%
}

\usepackage{longtable}

\begin{document}

\title{Navigating the Data Trading Crossroads: \\ An Interdisciplinary Survey}

\author{
Yi Yu$^{1}$, Jingru Yu$^{1}$, Xuhong Wang$^{1}$, Juanjuan Li$^{2}$, Yilun Lin$^{1,*}$, \textit{Member, IEEE} \\
Conghui He$^{1}$, Yanqing Yang$^{1}$, Yu Qiao$^{1}$, Li Li$^{3}$, \textit{Fellow, IEEE}, and Fei-Yue Wang$^{2,4}$, \textit{Fellow, IEEE}
\thanks{This work is supported by Shanghai Artificial Intelligence Lab}
\thanks{* for the corresponding author}
\thanks{$^{1}$Yi Yu (yuyi@pjlab.org.cn), Jingru Yu (yujingru@pjlab.org.cn), Xuhong Wang (wangxuhong@pjlab.org.cn), Yilun Lin (linyilun@pjlab.org.cn), Conghui He (heconghui@pjlab.org.cn), Yanqing Yang (yangyanqing@pjlab.org.cn), and Yu Qiao (qiaoyu@pjlab.org.cn) are with Shanghai Artificial Intelligence Laboratory, Shanghai 200240, China.}
\thanks{$^{2}$Juanjuan Li (juanjuan.li@ia.ac.cn) and Fei-Yue Wang (feiyue.wang@ia.ac.cn) are with the Institute of Automation, Chinese Academy of Sciences, Beijing 100190, China.}
\thanks{$^{3}$ Li Li (li-li@mail.tsinghua.edu.cn) is with the Department of Automation, BNRist, Tsinghua University, Beijing 100084, China}
\thanks{$^{4}$Fei-Yue Wang is also with the Institute of Engineering, Macau University of Science and Technology, Macau, China}
}

\markboth{Journal of \LaTeX\ Class Files,~Vol.~14, No.~8, August~2021}%
{Shell \MakeLowercase{\textit{et al.}}: A Sample Article Using IEEEtran.cls for IEEE Journals}

\maketitle

\begin{abstract}
Data has been increasingly recognized as a critical factor in the future economy. However, constructing an efficient data trading market faces challenges such as privacy breaches, data monopolies, and misuse. Despite numerous studies proposing algorithms to protect privacy and methods for pricing data, a comprehensive understanding of these issues and systemic solutions remain elusive. This paper provides an extensive review and evaluation of data trading research, aiming to identify existing problems, research gaps, and propose potential solutions. We categorize the challenges into three main areas: Compliance Challenges, Collateral Consequences, and Costly Transactions (the "3C problems"), all stemming from ambiguity in data rights. Through a quantitative analysis of the literature, we observe a paradigm shift from isolated solutions to integrated approaches. Addressing the unresolved issue of right ambiguity, we introduce the novel concept of "data usufruct," which allows individuals to use and benefit from data they do not own. This concept helps reframe data as a more conventional factor of production and aligns it with established economic theories, paving the way for a comprehensive framework of research theories, technical tools, and platforms. We hope this survey provides valuable insights and guidance for researchers, practitioners, and policymakers, thereby contributing to digital economy advancements.
\end{abstract}


\section{Introduction}
Information exchange has always been a crucial step in production and life throughout human history, driving cultural and scientific advancements. In antiquity, people organized hunts and shared experiences through oral communication. Ancient civilizations, such as Egypt and China, used papyrus and bamboo slips to record and transmit information. In more recent times, modern printing and postal systems have further facilitated communication. Entering the Digital Age, information exchange is primarily represented as data exchange, and data become a valuable factor of production. Reports show that the global data trading market reached \$ 311.72 billion in 2023, with an expected growth ratio at 14.9\% per year to attain USD 1088.06 billion by 2032 \cite{_global_}. Data showcases its significant value in various fields, particularly in artificial intelligence \cite{yuRoWbasedParallelControl2023,yuIdentifyingTrafficClusters2022,yu_pursuing_2023}. For instance, in recommendation systems, precise analysis of user data could provide personalized recommendations, enhancing user experience and boosting business profitability \cite{zhangStockMarketReactions2021}. The continuous advancement of artificial intelligence technology, specifically the development of large language models (LLMs), is fueling the exponential increase in demand for high-quality training data.

However, the increased demand for data has not resulted in a well-established market. Instead, data monopolies and privacy leakage emerged, hindering the development of the data market. Data monopolies arise when a few oligarchs control vast data resources, leveraging their significant financial capital and technological proficiency to govern and exploit them. Such situations significantly impede the circulation of data, hinder its effective utilization, and restrict the advancement of new technologies. Furthermore, misuse prevails in data trading among oligarchs. Without adequate legal and institutional safeguards for privacy and intellectual property rights, data collection, use, and transfer frequently leads to personal information leakage, causing widespread social concern and controversy. The phenomena of surging data demand coexist with data monopoly and misuse attracting increasing research interest. Most previous works focus on three problems:

Firstly, the Compliance Challenge arose from the lack of regulatory frameworks, industry standards, and governance mechanisms, causing data trading to be accompanied by significant uncertainty and leading to prominent concerns regarding data monopolization and misuse. The compliance challenge is fundamentally caused by the limited understanding of data and its rights attribution. As a new factor of production with non-exclusivity, dynamism, and privacy characteristics, data's value evaluation, trading mechanism, and social impact are still under study, thereby causing a lag in institutional development \cite{peiDataPricingData2021}.

Secondly, there exist Collateral Consequences during data circulation, manifested in the widespread social concern and controversy of data trading since the external effects are not sufficiently internalized. For example, trading personal data could cause potential privacy leakage in data trading, which can significantly impact individuals and even society, but such impacts are not considered or compensated. The collateral consequences arise primarily due to the unclear responsibility during circulation, which implies no one is responsible for social implications caused by the non-rivalrous, private, and indivisibility characteristics of the data \cite{yu_swdpm_2023}.

Thirdly, a Costly Transaction problem exists during data trading, which manifests in various costs in negotiation, contracting, and enforcement. For example, data's dynamic and indivisible characteristics require preserving its "freshness" throughout trading, requiring increased infrastructure investments to ensure frequent updates, processing, and transmission. As a result, an escalation in infrastructure leads to increased implicit costs and high trading thresholds, suppressing the activity of the data market \cite{peiSurveyDataPricing2020}. Fundamentally, the Ambiguity of cost-sharing during data trading hinders the trading process, impeding the establishment of a cohesive and thriving market despite the substantial demand for data.

\begin{figure} [htbp]
    \centering
    \includegraphics[width=0.6\linewidth]{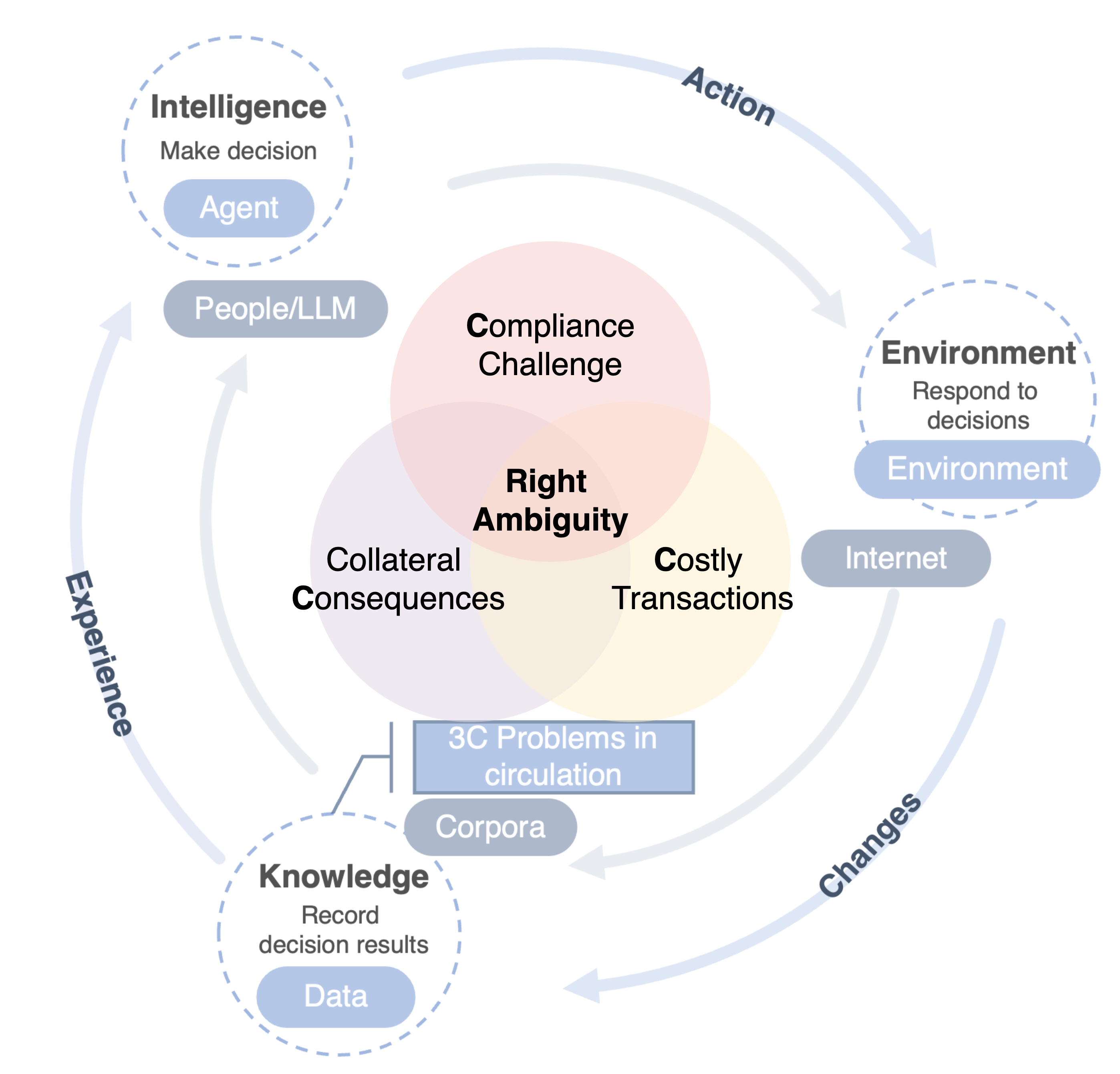}
    \caption{The 3C problems in data trading that interrupts the intelligence-environment-data circulation.}
    \label{fig:problems}
\end{figure}

As illustrated in Fig.\ref{fig:problems}, we summarize the aforementioned problems as the 3C problems (Compliance Challenge, Collateral Consequences, and Costly Transactions), all pointing to one fundamental obstacle in data circulation: Right Ambiguity. Right Ambiguity refers to unclear definitions of data's property rights, entitlements, interests, and responsibilities. Right Ambiguity poses obstacles to effectively utilizing data's economic attributes, thereby influencing the legality, compliance, and operability of data trading. Furthermore, it disrupts the current flow of intelligence-environment-data circulation, which in turn impedes intelligence development and further hampers the formation of a healthy and active market. 

To address the 3C problems, academia and industry have put forth various methods, such as establishing regulatory frameworks, adopting privacy protection technologies during trading, and proposing data pricing methods. Research is advancing towards more integrated and systematic solutions, driven by the emergence of cutting-edge technologies such as blockchain and smart contracts. These solutions are not restricted to discussing a single technology; instead, they emphasize the importance of integration to address the 3C problems, such as constructing decentralized data markets that cope with the complexity and diverse needs of data trading via blockchain technology \cite{li_multiblockchain_2023}. 
Several surveys have reviewed these data trading studies. Some analyzed technological innovations from the perspective of the data lifecycle \cite{liang_survey_2018}, while others explored the status and methods of data trading from economic perspectives\cite{peiSurveyDataPricing2020} and legal \cite{asswadDataOwnershipSurvey2021} perspectives. These surveys provide a detailed analysis of technological advancements and serve as valuable references for research and practical applications. However, these studies neglect the fundamental obstacle, Right Ambiguity, in data trading.

In this paper, we present a comprehensive analysis of the 3C problems encountered in data trading. We systematically review existing methods and solutions developed to tackle these problems. Our findings suggest the Right Ambiguity, which hinders data from being traded like a mature commodity, eventually causing 3C problems. Building upon this insight, we propose data usufruct, which refers to the right of non-data owners to use and benefit from data, as a systemic solution. By separating data usufruct via privacy-preserving computing, identity authentication, and fingerprinting technologies, data can be transformed into a common factor of production, aligning with existing economic theories. We further propose a data usufruct research framework for addressing the 3C problems. 

It is crucial to highlight that future directions should involve the integration of diverse disciplines, including economics, sociology, and computer science, to establish a comprehensive research framework for data usufruct. The framework should encompass theories, tools, and platform development for trading data usufruct via leveraging advanced technologies. These endeavors are anticipated to address the three key challenges (3C problems) while fostering the sound growth of the data trading market.

The key contributions of this paper can be summarized as:
(1) revealing the 3C problems and right ambiguity as the fundamental obstacle in data trading; (2) conducting a comprehensive review of data trading studies from the perspective of solving 3C problems; (3) proposing the data usufruct as a systemic solution to address 3C problems, and suggesting future research direction.

\section{Quantitative Analysis of previous works}
In this section, we perform quantitative analyses on a dataset comprising 6,865 articles pertaining to the subject of data trading. These articles were obtained by searching in the Web of Science database using the keyword "data trading" and filtering for English papers published after 2000 within the Web of Science Core Collection. Through these analyses, we get an overview of the 
research trajectory transitioning from single-point methods to integrated solutions.

\subsection{Publication Year Distribution Analysis}
As depicted in Fig. \ref{fig:record_count}, the number of publications has steadily increased since 2000, with a notable surge after 2017. Specifically, there were only 47 publications in 2000; however, by 2023, this count had escalated to 766, indicating substantial research interests in data trading.

\subsection{Geographical Distribution Analysis}
Fig. \ref{fig:country_analysis} illustrates the geographical distribution of the literature, revealing a concentration of research centers in China and the United States, accounting for over 60\% of the total research output. These two countries emerge as primary contributors to research in this field, while other European nations such as Germany and the United Kingdom also contribute relatively high research volume.

\subsection{Research Area Distribution Analysis}
\begin{figure}[!t]
\centering
\subfloat[]{\includegraphics[width=2.5in]{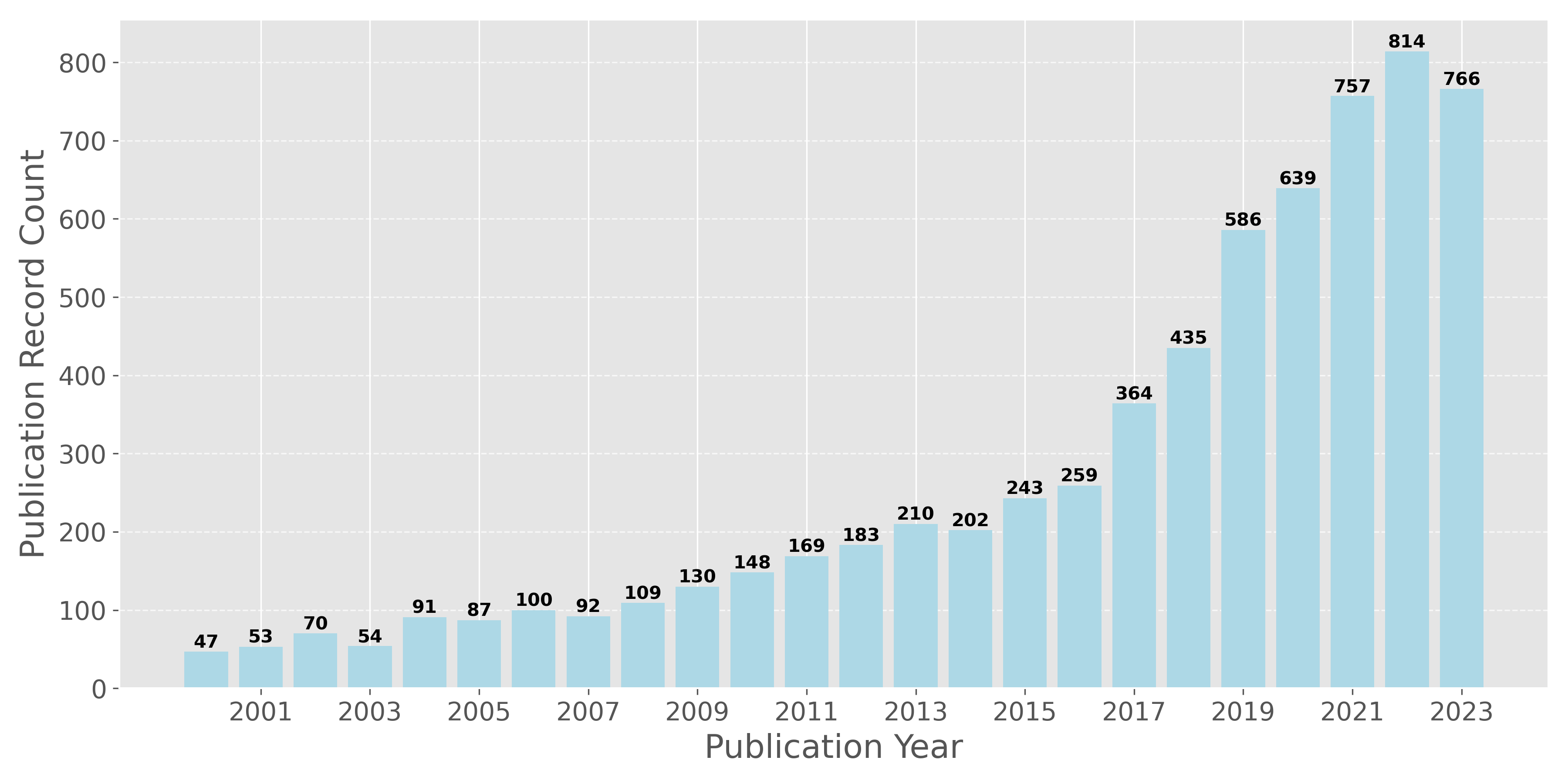}%
\label{fig:record_count}}
\hfil
\subfloat[]{\includegraphics[width=0.5\linewidth]{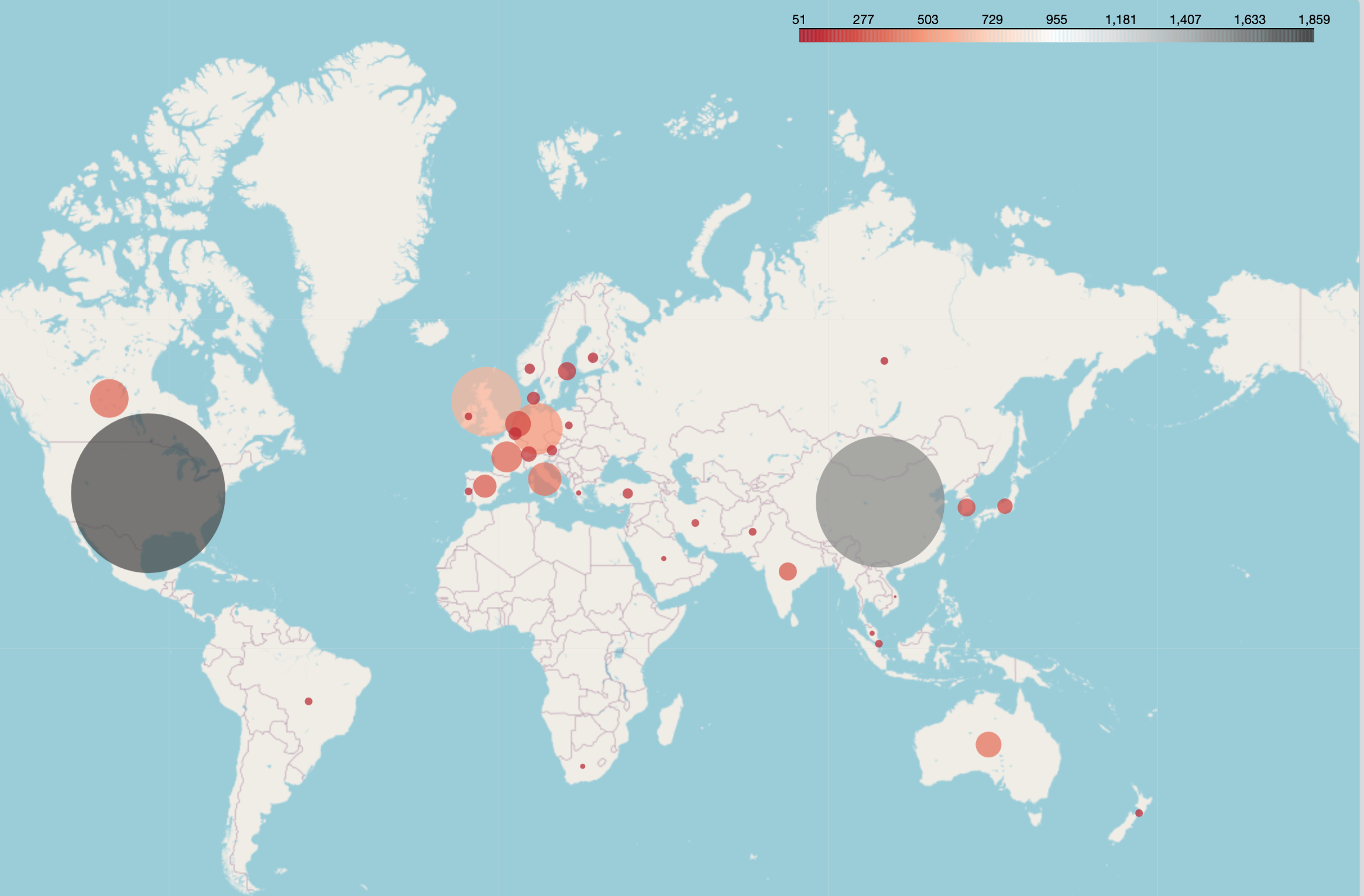}%
\label{fig:country_analysis}}
\subfloat[]{\includegraphics[width=0.5\linewidth]{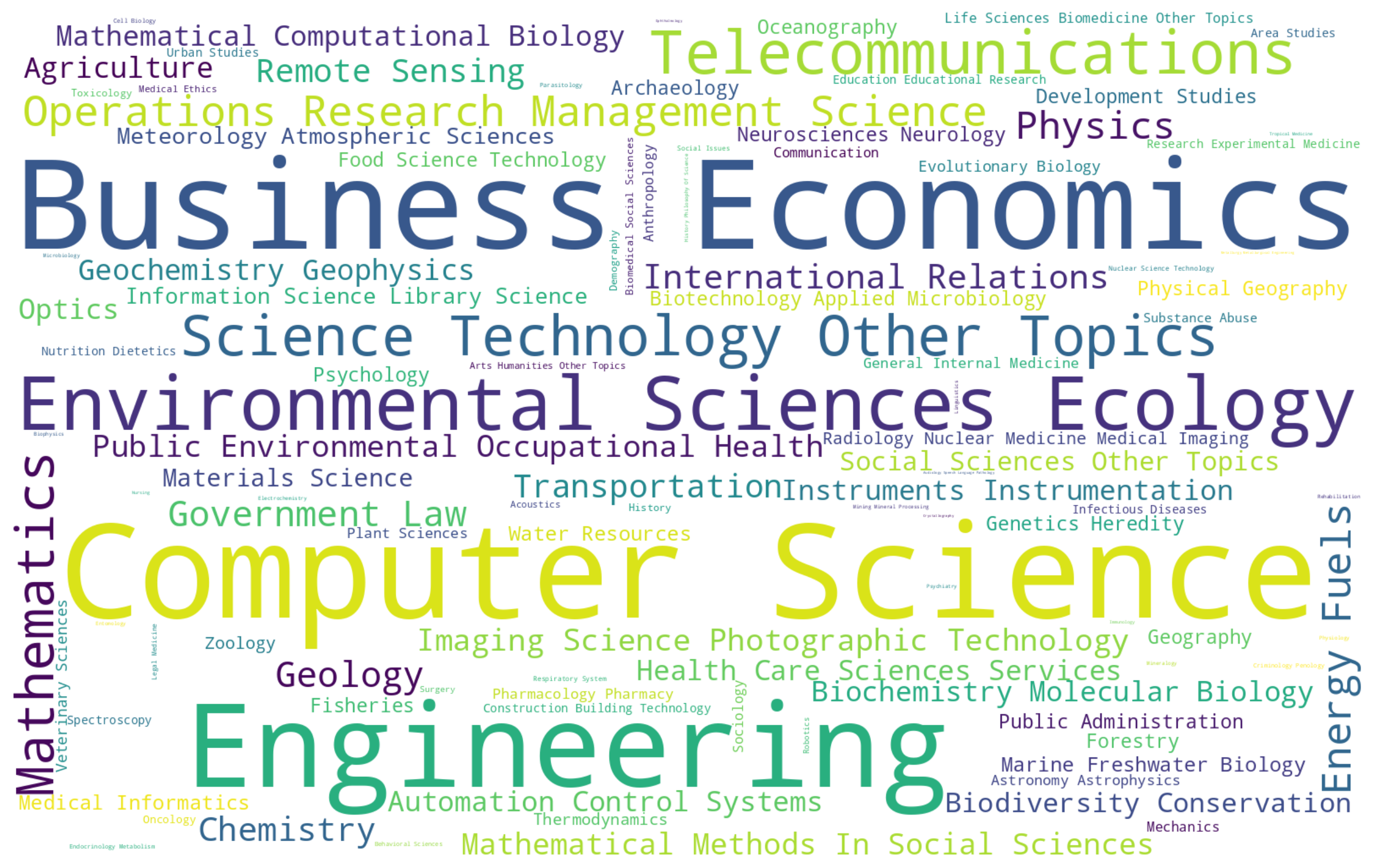}%
\label{fig:research_areas}}
\caption{Distribution analysis of literature since 2000. (a) Explosive growth in data trading publication record count. (b) Geographical distribution of publications, showing predominance in the United States and China.(c) Word-cloud analysis of research fields, indicating major focus on computer science, business economics, and engineering}
\label{fig:year_analysis}
\end{figure}

As shown in Fig. \ref{fig:research_areas}, data trading receives the most attention in the fields of computer science, business economics, and engineering, with research proportions of 24.4\%, 22.7\%, and 19.7\%, respectively. In addition, a considerable proportion of research is conducted in environmental sciences and telecommunications.

The analysis above demonstrates a significant rise in research on data trading since 2000, with explosive growth in recent years. The focus of research on data trading is centered primarily in the United States and China, with computer science, business economics, and engineering emerging as the dominant fields. This indicates that data trading is gaining increasing attention as an interdisciplinary research topic.

\subsection{Keyword Co-occurrence Analysis}

\begin{figure}[htb]
    \centering
    \includegraphics[width=1\linewidth]{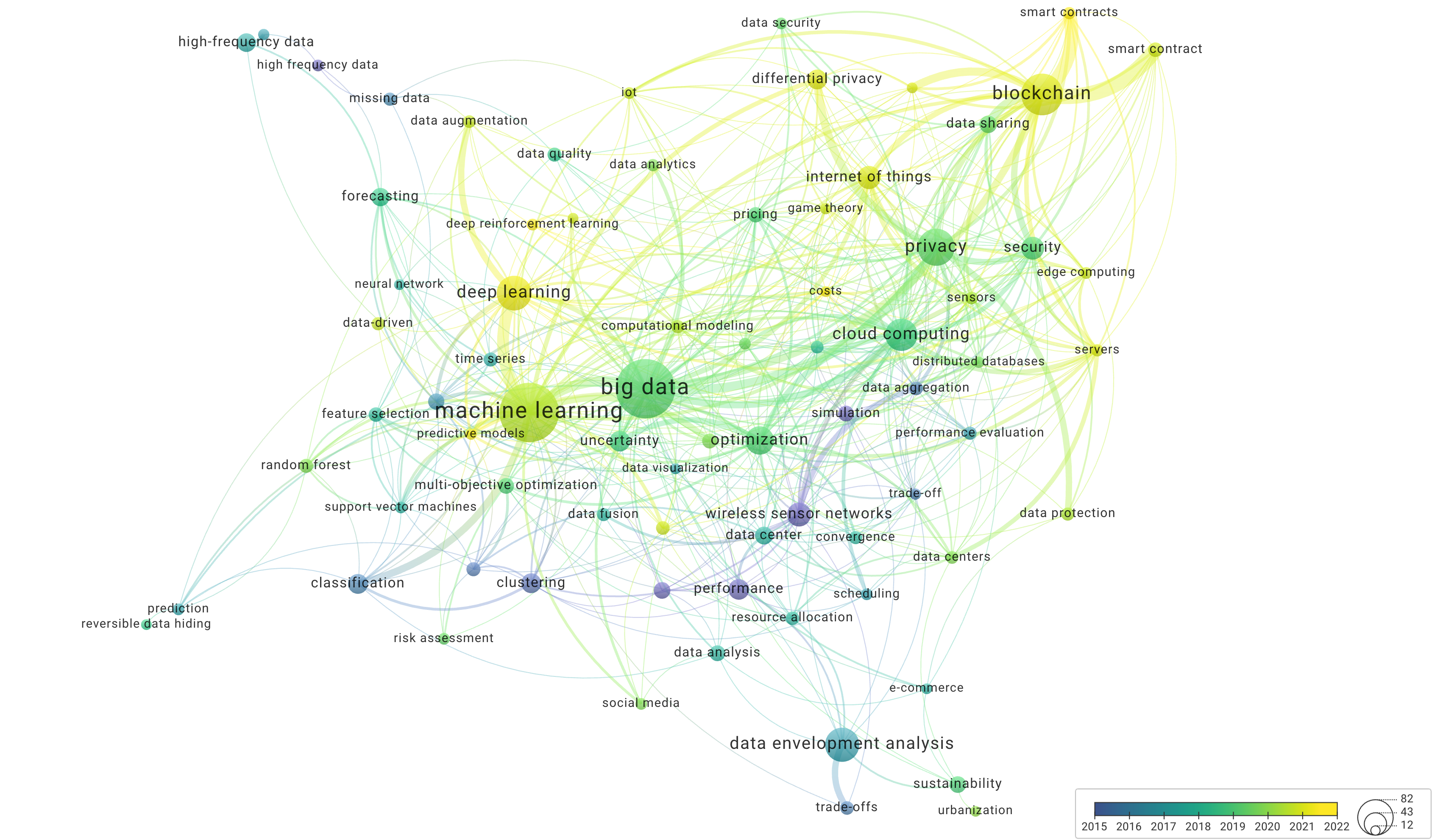}
    \caption{Keyword co-occurrence analysis in data trading research since 2000: evolving from isolated methods to systemic solutions (Source: Web of Science)}
    \label{fig:research_trend}
\end{figure}

We further perform a co-occurrence analysis of keywords in the literature using the VOSviewer\cite{vaneck_software_2010}, as depicted in Fig. \ref{fig:research_trend} and Table \ref{tab:keyword_analysis}. These visualizations illustrate the correlation between research hotspots and temporal transitions in research focus.

\begin{table}[h]
\centering
\caption{Analysis of Key Technologies}
\label{tab:keyword_analysis}
\begin{tabular}{@{}p{3.5cm}p{2cm}p{1cm}p{1cm}@{}}
\toprule
Technology & Average Publication Year & Occurrence Count & Average Citation Count \\ \midrule
Data Envelopment Analysis & 2017.1  & 78  & 21.60  \\
Cloud Computing           & 2019.3  & 72  & 30.97  \\
Big Data                  & 2019.7  & 172 & 21.55  \\
Privacy                   & 2019.9  & 88  & 15.69  \\
Machine Learning          & 2021.0  & 172 & 18.91  \\
Blockchain                & 2021.5  & 104 & 24.63  \\ \bottomrule
\end{tabular}
\end{table}

The keywords in Fig. \ref{fig:research_trend} exhibit a transition trend, which can be categorized into three distinct clusters: blue-purple, green, and yellow:

\begin{itemize}
\item Blue-purple: \textbf{Data Envelopment Analysis} was applied earlier in data trading research, mainly used for efficiency evaluation and optimization. \textbf{Clustering} algorithms are used to model data distribution characteristics, improving the efficiency of data analysis and processing.
\item Green: \textbf{Big Data} was used for data storage and processing in the early stage (before 2015). Over time, the integration of machine learning with Big Data technology has enabled advanced data analysis and prediction. The combination of \textbf{Cloud computing} and big data technology has gained substantial momentum since 2017 due to its widespread application in data storage, processing, and analysis. \textbf{Privacy} computing technology has attracted significant attention since 2019 as privacy concerns have grown.
\item Yellow: \textbf{Blockchain} technology is closely related to privacy protection, especially after 2021, mainly associated with keywords such as \textbf{smart contracts}, \textbf{differential privacy}, and \textbf{game theory}, showing an integrated research trend in data trading.
\end{itemize}

Based on the literature analysis, it is evident that the research trend in data trading has evolved from single-point approaches to systemic solutions. Prior to 2020 (as depicted by the blue and green clusters in Fig. \ref{fig:research_trend}), research primarily focused on isolated breakthroughs. For compliance challenges, various techniques such as data envelopment analysis, cloud computing, and big data analysis have been employed to improve processing and storage efficiency while serving as a robust data circulation system. Privacy-preserving computing technologies are utilized to alleviate potential collateral consequences. To tackle costly transaction issues, methods for data evaluation and pricing are implemented. Moreover, recent works have shifted towards more systemic and integrated solutions, such as integrating machine learning with blockchain technology. This integration has resulted in significant advancements in tackling 3C problems through blockchain.

\section{Evolution of Research in Data Trading}
\begin{figure}[!t]
\centering
\includegraphics[width=0.8\linewidth]{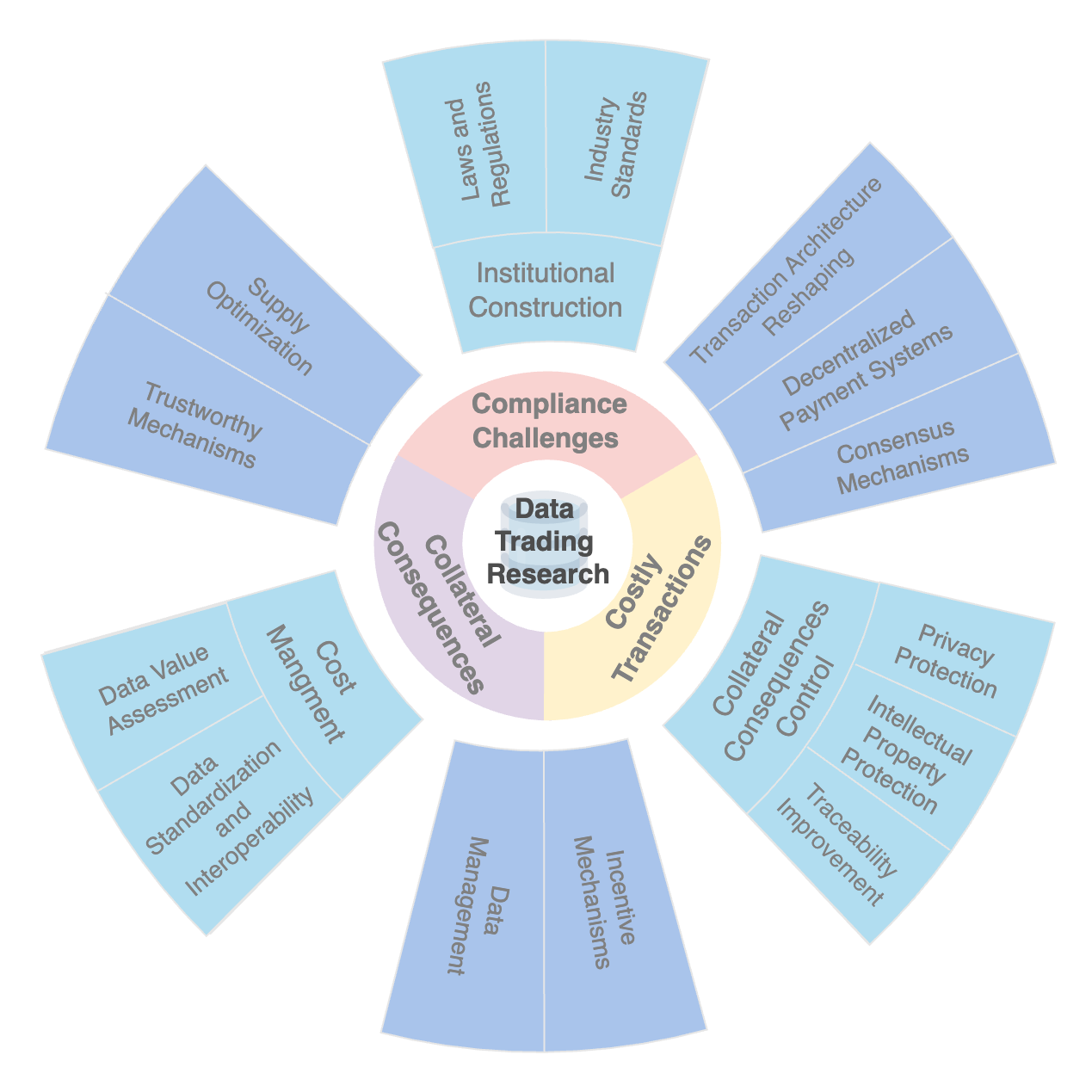}
\caption{Review framework of data trading research: from single-point methods to systemic solutions}
\label{fig:review_framework}
\end{figure}

The literature review framework in this section is illustrated in Fig. \ref{fig:review_framework}, which is derived from the evolving research trajectory and technologies. The review is divided into two parts: an exploration of research conducted using single-point methods, followed by an evaluation of integrated solutions. In this section, we aim to enhance the understanding of the current research landscape through a systematic literature review while offering comprehensive guidance for academia and industry.

\subsection{Single-point methods}
\subsubsection{Institutional Construction}

From a legal perspective, there is a fundamental conflict between protecting data privacy and data trading, which significantly increases the complexity of data usage and sales \cite{Reimers2023FromDE}. To create an effective data trading system, finding a delicate balance between ensuring robust privacy protection and facilitating data circulation is imperative. Scholars have proposed several methods to address these compliance challenges, including revising existing laws and regulations, formulating industry standards, and establishing comprehensive regulatory mechanisms. These methods aim to build a fair and transparent data trading infrastructure, reducing uncertainty and risk for all trading participants.


\paragraph{Laws and Regulations}

Various proposals have been implemented in data trading institutional construction, involving data property rights confirmation, the "three rights separation" system, hierarchical authorization system, and international cooperation \cite{gao_data_2020,gong_research_2019,liu_docs_2022,_china_,xu_bhda_2020,xuan_hierarchically_2019,shabani_data_2021,reidenberg_resolving_2000}. \citeauthor{paprica_essential_2023} propose measures to manage the basic norms of data users by formulating laws and systems, as well as strengthen public participation and safeguard data sovereignty, providing clear legal basis and protection for data owners, users, and traders \cite{paprica_essential_2023}.

To address the ambiguity of data ownership, researchers have investigated implementing a registration system for data property rights. Data property rights registration is similar to real estate registration, used to clarify data ownership \cite{gao_data_2020,gong_research_2019,liu_docs_2022}. \citeauthor{gao_data_2020} propose a data right confirmation mechanism based on blockchain and locality-sensitive hashing, which not only clarifies data ownership but also lays the foundation for the legal and orderly circulation of data \cite{gao_data_2020}. Similarly, \citeauthor{gong_research_2019} and \citeauthor{liu_docs_2022} leverage smart contracts and blockchain to ensure the auditability, accountability, and integrity of data trading while confirming data ownership \cite{gong_research_2019,liu_docs_2022}.

In response to the complexity of data ownership, China unveils 20 key measures to build basic systems for data, innovatively proposes a data property rights mechanism framework that covers data resource ownership, data processing, usage rights, and management rights of data-related products\cite{_china_}. At the same time, by exploring the hierarchical authorization system for data classification and grading, data is classified and graded according to its importance and sensitivity, and corresponding permissions are granted to different entities according to different categories and levels \cite{xu_bhda_2020,xuan_hierarchically_2019}. \citeauthor{xu_bhda_2020} propose a blockchain-based hierarchical data access model for financial services, which contains consent management and dynamic credits management, implementing fine-grained data access control through rating accredited data recipients (ADRs) \cite{xu_bhda_2020}. These approaches contribute to clarifying the responsibilities and rights about different data entities, facilitating efficient utilization of data, and ensuring organized data circulation.

Facing the international data trading challenges, there are some data trading regulatory acts worldwide. The European Union(EU) Data Governance Act promotes data sharing through good interaction with the General Data Protection Regulation (GDPR) \cite{shabani_data_2021}. \citeauthor{shabani_data_2021} discuss how the Data Governance Act interplays with the GDPR and its potential impact on biomedical research, emphasizing the need for further improvements to impact data markets positively \cite{shabani_data_2021}. In addition, the EU is working to formulate legislation for data \cite{riis_shaping_2022}. \citeauthor{riis_shaping_2022} demonstrate how EU data law can be classified as an autonomous legal field and diverges from adjacent legal fields by striving to safeguard five distinct objectives: protection of a competitive market, fundamental rights, consumers, trustworthiness, and open data \cite{riis_shaping_2022}. At the same time, countries participating in the Comprehensive and Progressive Agreement for Trans-Pacific Partnership (CPTPP) and the United States-Mexico-Canada Agreement (USMCA) are also actively promoting the legal construction of data circulation and building a digital trade regulatory model \cite{burri_data_2021, burri_governance_2017}. \citeauthor{burri_data_2021} trace the developments in preferential trade agreements, particularly those of relevance to cross-border data circulations, and provide a detailed analysis of the current state of global digital trade law \cite{burri_data_2021}. The development of the CPTPP and USMCA demonstrates how to reflect the increasing importance of data circulation through trade law, as well as address the difficulty of controlling data circulation to protect critical public interests \cite{burri_governance_2017}. 

Moreover, international cooperation is crucial for resolving legal conflicts in cross-border data circulation, given significant policy and regulatory discrepancies across countries concerning data privacy protection. 
In international cooperation, by establishing a common information privacy core principle framework and conducting co-regulation, confrontation arising from governance choices can be avoided \cite{reidenberg_resolving_2000}. \citeauthor{reidenberg_resolving_2000} explore the divergences in approach and substance of data privacy between Europe and the United States, and offer a conceptual framework for coregulation of information privacy that can avoid confrontations over governance choices \cite{reidenberg_resolving_2000}. The role of international trade law in internet governance, especially the framework construction to promote data circulation by balancing domestic Internet regulation and liberalized data circulation, also indicates its potential in promoting data circulation \cite{mishra_building_2018}.

Establishing laws and regulations provides a robust institutional framework for data trading. However, to ensure the adequate circulation and utilization of data, it is also crucial to develop industry standards at a technical level\cite{fagin_data_2003, suzuki_datatrading_2021}.

\paragraph{Industry Standards}

The formulation of industry standards addresses concerns regarding the efficiency and safety of data circulation. By establishing unified frameworks for technical specifications, data formats, and trading processes, industry standards enhance the transparency and efficiency of data trading while safeguarding the privacy and intellectual property rights of individuals. In particular, standards play a crucial role in overseeing the quality and security of data, which is essential for ensuring the reliability and integrity of traded information. Furthermore, industry standards facilitate interoperability between different platforms and systems, reducing technical barriers and supporting broader circulation and utilization of data.

Additional conversions may be required during data trading to maximize the effectiveness of data obtained from various sources. \citeauthor{arenas_data_2013} propose efficient data exchange and conversion mechanisms to ensure data consistency and accuracy \cite{arenas_data_2013}. \citeauthor{fagin_data_2003} identify a particular class of solutions called universal solutions in data exchange and show that the core of a universal solution is also a universal solution, and hence the smallest universal solution \cite{fagin_data_2003}. \citeauthor{suzuki_datatrading_2021} standards for data encryption, anonymization, and interfaces enhance data security and interoperability, further contributing to the efficiency and value of data trading \cite{suzuki_datatrading_2021}. These studies provide an essential theoretical and practical basis for formulating industry standards for data trading.

Enhancements in legal frameworks and the establishment of technical standards significantly improve the institutional environment for data trading, thereby facilitating the high-quality development of the data market\cite{gao_data_2020,gong_research_2019,liu_docs_2022,_china_,xu_bhda_2020,xuan_hierarchically_2019,reidenberg_resolving_2000, arenas_data_2013, fagin_data_2003, suzuki_datatrading_2021}. 

\subsubsection{Collateral Consequences Control}

Nevertheless, effectively addressing collateral consequences such as data misuse and privacy infringement continues to pose a challenge, necessitating additional oversight measures. The non-excludable and non-rivalrous characteristics of data may lead to its over-exploitation, meanwhile, the sensitivity of data may lead to misuse, both of which can have negative collateral consequences. To mitigate these consequences effectively, various measures such as privacy computing, data watermarking, and tracking are implemented for efficient data management. 

\paragraph{Privacy Protection}

Data trading, which may involve personal information and potentially impact national security, requires additional methods to mitigate substantial collateral consequences. This issue has gained widespread attention as a prominent concern within the industry \cite{yu_swdpm_2023}. However, there exists an inherent tension between safeguarding data privacy and maximizing value extraction: placing excessive emphasis on privacy protection may undermine the potential value of data, whereas misuse of data can infringe upon individuals' privacy.

Privacy-preserving computing techniques such as encryption and anonymization are implemented to achieve data available but not visible at the technical level, safeguarding data security and privacy, offering new insights for addressing the dichotomy between data privacy and value realization \cite{ouyang_artificial_2023}.

Differential privacy, for instance, protects personal privacy by adding noise to published data to obscure individual identities, which is applicable in various data-sharing scenarios\cite{olaleye_composition_2022}. Techniques for implementing differential privacy vary to accommodate different needs and ensure a balance between privacy protection and data utility \cite{kellaris_practical_2013, xiong_effective_2020}. \citeauthor{kellaris_practical_2013} propose a Grouping and Smoothing (GS) method that can reduce data perturbation while maintaining $\epsilon$-differential privacy when publishing statistical data, which is beneficial for data trading in application scenarios such as traffic and medical data \cite{kellaris_practical_2013}. In addition, \citeauthor{xiong_effective_2020} prove that using Local Differential Privacy (LDP) techniques to perturb data at the user side is particularly effective in health application data collection settings, providing stronger privacy guarantees than Differential Privacy (DP) \cite{xiong_effective_2020}. Their experiments also show that when shown descriptions that explain the implications instead of the definition/processes of DP or LDP technique, participants demonstrated better comprehension and showed more willingness to share information with LDP than with DP, indicating their understanding of LDP's stronger privacy guarantee compared with DP. In practical data trading, applying differential privacy techniques requires a comprehensive consideration of privacy protection, data sharing requirements, and fairness issues to achieve an effective balance between privacy protection and data sharing.

Encryption-based privacy protection methods are also introduced into data trading research. \citeauthor{xue_blockchainbased_2023} propose a blockchain-based data trading scheme that uses attribute-based encryption and zero-knowledge proofs to ensure the fairness and privacy of transactions \cite{xue_blockchainbased_2023}. \citeauthor{zhao_machine_2019} propose a new blockchain-based data trading protocol for big data markets, using ring signatures and double-authentication-preventing signature systems to protect the privacy of data providers \cite{zhao_machine_2019}. The suggested protocol combines ring signature, double-authentication-preventing signature, and similarity learning to ensure the availability of trading data, privacy for data providers, and fairness between data providers and consumers. Authors define similarity on specified data and learn a distance function to conduct efficient retrieval, and the data consumer uses similarity learning to decide whether to purchase the data or not. \citeauthor{gao_privacypreserving_2020} propose a privacy-preserving auction scheme using homomorphic encryption to enhance the security of data auctions \cite{gao_privacypreserving_2020}. \citeauthor{garrido_revealing_2022} systematically reviews the application of privacy-enhancing technologies in IoT data markets and points out the limitations of existing technologies and future research directions \cite{garrido_revealing_2022}. Their findings suggest that there is no established canonical architecture for an IoT data market. Current solutions often lack the necessary combination of anonymization and secure computation technologies.

While privacy computing has advanced significantly, the implementation of these services in real-world data trading is often delayed. This delay primarily stems from a greater emphasis on safeguarding the data assets of influential entities rather than ensuring comprehensive protection of personal privacy. Consequently, preserving privacy continues to be a challenge in data trading \cite{garrido_revealing_2022}.

\paragraph{Intellectual Property Protection}
As a replicable asset, data faces the challenge of unauthorized disclosure and illegal copying during circulation. Watermarking and fingerprinting technologies have emerged to provide new solutions to this problem. Before trading or utilizing data, it is possible to invisibly embed watermark information into the data, thereby indicating the copyright of the data. Once data leakage is detected, the data's ownership and the violating party can be determined through watermark extraction and verification, ensuring the integrity and traceability of the data and providing a basis for rights protection \cite{olaleye_composition_2022}. A unique digital fingerprint can be generated, which enables efficient identification of the data source, tracking of its propagation trajectory, and protection of its ownership \cite{wu_privacyfriendly_2019,sheng_cpchain_2020,li_achieving_2022}. \citeauthor{sheng_cpchain_2020} presents a blockchain-based framework called CPchain that aims to preserve copyright in crowdsourcing data trading \cite{sheng_cpchain_2020}. This framework incorporates ring signature, double-authentication-preventing signature, and similarity learning techniques to ensure data accuracy while maintaining individual rationality and quality of data. CPchain can also combine digital fingerprint technology with blockchain to protect data copyright without a third-party certification authority. \citeauthor{li_achieving_2022} propose a fair and accountable data trading scheme for educational multimedia data based on blockchain, which combines anti-collusion code named BIBD-ACC with asymmetric fingerprinting technology, aiming to achieve robust copyright protection\cite{li_achieving_2022}.

\paragraph{Traceability Improvement}

Data traceability is crucial for controlling the collateral consequences of data circulation. Data abuse and unauthorized use can be effectively prevented by recording and tracking the circulation trajectory of data, promoting the development of "fair trade" data \cite{groth_transparency_2013}. \citeauthor{sheng_cpchain_2020} proposes a full-chain copyright protection scheme combining data watermarking and fingerprinting technologies to effectively mitigate infringement in data trading and create an orderly trading environment \cite{sheng_cpchain_2020}. Data fingerprints can be used to judge the accuracy of the data, thereby determining the truthfulness of data \cite{yang_blockchainbased_2020,zhang_privacypreserved_2024a}. \citeauthor{zhang_privacypreserved_2024a} propose a privacy-preserved data disturbance and truthfulness verification scheme for data trading, incorporating a private-verifiable imprint-embedded disturbance method and implementing adaptive truthfulness verification \cite{zhang_privacypreserved_2024a}.

Constructing auditing mechanisms can also mitigate the collateral consequences of data circulation. Full lifecycle auditing focuses on data security and compliance throughout the entire process from generation to destruction \cite{olaleye_composition_2022}. By recording the trajectories of data and data usage via blockchain, it is possible to monitor data circulation and prevent unauthorized use and abuse.

Efforts in privacy protection, intellectual property preservation, and data traceability contribute to controlling the collateral consequences of data circulation. This creates a conducive environment for the robust growth of the data market. 

\subsubsection{Cost Management}

Apart from institutional development and management of collateral consequences, reducing costs in data trading also plays a crucial role in fostering the advancement of the data market. Reports show that approximately 17\% of large public and private companies participate in the data market, and most data transactions (80\%) involve customer data. However, more than half of data transactions adopt mutual data-sharing agreements rather than mature commodity sales models \cite{elsaify_data_2020}, reflecting low market activity and high data trading costs.

Data trading costs encompass explicit expenses such as information acquisition, processing, storage, and transmission, as well as implicit expenses arising from information asymmetry and other factors. To mitigate the burden of high trading costs, existing approaches concentrate on streamlining operational processes of data trading and reducing trading thresholds. Standardized trading processes are established to diminish operational complexity and information asymmetry during transactions. Additionally, efforts are made to devise more precise and dynamic data pricing mechanisms that accurately reflect actual data value and foster equitable market transactions.


\paragraph{Data Standardization and Interoperability}

Data from different sources vary significantly in format, semantics, quality, and other aspects. Lacking unified standards and norms leads to high data interconnection, integration, and application costs, making it difficult to unleash the value of data fully. Data format and interface standardization are vital to achieving data interconnection \cite{elsaify_data_2020}. This reflects a unique characteristic of data, namely heterogeneity. Heterogeneity results in elevated costs in data interoperability, hindering the large-scale development of data trading. The solution to this problem lies in the establishment of mechanisms for data standardization.

The challenges arising from data heterogeneity can be mitigated effectively by standardizing elements such as data formats, models, and interface protocols and applying key technologies such as semantic mapping, data cleaning, and data synchronization \cite{singh_crossdomain_2021,koshizuka_dataex_2022,iwasa_development_2020}. \citeauthor{koshizuka_dataex_2022} introduce DATA-EX, an infrastructure designed for cross-domain data exchange using a federated architecture\cite{koshizuka_dataex_2022}. This framework establishes open standard specifications and guidelines for data exchange and platform operation. Additionally, it includes software systems based on these standards and governance rules to ensure the smooth operation of DATA-EX. \citeauthor{iwasa_development_2020} develop a new online platform for stakeholders to communicate regarding data utilization, which is superior to the conventional paper-based IMDJ in terms of reducing the burden on conducting workshops and enabling cross-domain data exchange and cooperation \cite{iwasa_development_2020}. By promoting interoperability and standardization, it is possible to eliminate barriers to data circulation at the system architecture level, connecting data silos.

\paragraph{Data Value Assessment}

Data pricing is one of the core issues in building the data market. Unlike traditional commodities, data is difficult to measure in value due to the absence of standardized pricing criteria and regulations. Consequently, it results in elevated costs for data trading. 

Researchers develop theoretical frameworks for data pricing by considering various dimensions, including data asset attributes, data quality, and supply and demand in the data market. Diversified data pricing models are developed by employing cost pricing, revenue pricing, market pricing, and other approaches \cite{shen_personal_2022,liao_establishing_2023,liang_data_2021}. \citeauthor{shen_personal_2022} presents a novel approach, known as the Personal Big Data Pricing Method (PMDP), based on Differential Privacy, which introduces two distinct pricing mechanisms: positive and reverse pricing \cite{shen_personal_2022}. In positive pricing, the compensation for privacy is initially computed based on the individual loss of privacy, and the query price is determined by total privacy compensation. In reverse pricing, the query price is first calculated and then a fair allocation of compensation is made according to the loss of privacy. These mechanisms aim to determine the appropriate prices for personal big data effectively. Results show that PMDP can provide reasonable pricing for personal big data and fair compensation to data owners, ensuring an arbitrage-free condition and finding a balance between privacy protection and data utility. 
\citeauthor{yu_swdpm_2023} propose a social welfare-optimized data pricing mechanism (SWDPM) in multi-round data trading scenarios. Markov decision process is used to model progressive information disclosure in multi-round trading, and SWDPM is used to find optimal strategies. Numerical results show that SWDPM could increase social welfare 3 times in data trading. \citeauthor{liao_establishing_2023} propose a utility function to quantify the privacy calculus mechanism for users, considering three choices: purchasing, selling, or maintaining\cite{liao_establishing_2023}. The model can solve the dilemma of enterprises violating privacy laws by profiting from co-creation data without user permission and realize a Pareto improvement of the utilities of the platform and users. \citeauthor{liang_data_2021} develop a comprehensive index system for data prices based on hedonic price theory, encompassing three key dimensions: data objects, data demanders, and data suppliers \cite{liang_data_2021}. The empirical analysis demonstrates that logarithmic functions can effectively capture the relationship between hedonic price models and various factors affecting data prices. Furthermore, they successfully identify significant correlations between these factors and the pricing of data products.

Intelligent data assessment tools are concurrently being developed by utilizing artificial intelligence, big data analysis, and other cutting-edge technologies. Data pricing and trading platforms are built to achieve automated and real-time data pricing, significantly reducing trading costs \cite{hayashiTEEDAInteractivePlatform2020}. \citeauthor{hayashiTEEDAInteractivePlatform2020} propose a new online platform called Web Innovators Marketplace on Data Jackets (Web IMDJ) for stakeholders to communicate regarding data utilization \cite{hayashiTEEDAInteractivePlatform2020}. Web IMDJ surpasses the traditional paper-based IMDJ by alleviating the workload associated with organizing workshops. Additionally, it facilitates seamless communication between data providers and consumers, fostering collaboration and enhancing data usage.

Despite the abundance of research on data pricing, current data markets still face acceptance challenges after evaluation. There is a pressing need for further exploration and enhancement of practical data pricing methods. It is essential to acknowledge that data pricing, trading payment, and incentives are interconnected and exert mutual influence on each other, collectively shaping the data trading ecosystem. It is challenging to address these three aspects in isolation as their synergistic effects necessitate a comprehensive consideration for optimizing the data trading market. The coordination of all aspects of data trading is expected to be explored further in the future. 

In summary, the existing methods attempt to address 3C problems in data trading by institutional development, collateral consequences control, and transaction cost management. The enhancement of laws and regulations and the establishment of industry standards provide institutional assurances for data trading. Measures such as safeguarding personal privacy, protecting intellectual property rights, and managing data traceability control the adverse effects resulting from data circulation. Additionally, efforts to standardize data, promote interoperability construction, and innovate value assessment mechanisms contribute to reducing trading costs. These collective endeavors effectively foster the growth of the data market. However, since 3C problems are intricate and intertwined, singularly addressing one problem cannot solve the problems globally. Numerous challenges persist in data trading, necessitating further investigation from diverse perspectives.

\subsection{Integrated solutions}

Due to the correlation between the compliance, externality, and cost problems in data trading, it is challenging to solve all problems individually. To address 3C problems, cutting-edge academic research has proposed solutions trying to simultaneously address multiple problems

\subsubsection{Addressing Compliance Challenge and Collateral Consequence}
The expansion of the data trading market is closely tied to the presence of well-established institutions, whereas their absence has resulted in severe collateral consequences. For example, the lack of trust in data trading hinders the assessment of trading parties' performance and commitment. Information asymmetry is also prevalent, impeding data market development. This scenario can be likened to the "Market for Lemons" phenomenon described by \citeauthor{akerlof_market_1978}, where quality uncertainty and information asymmetry lead to adverse selection, driving out high-quality data \cite{akerlof_market_1978}.

According to \citeauthor{carro_markets_2015}, imperfect institutions give rise to external information sources, such as advertising, public opinion, or rumors, influencing trader behavior and resulting in market bias \cite{carro_markets_2015}. \citeauthor{sefton_information_2001} emphasizes that as the number of uninformed consumers increases, market prices become less competitive for all consumers, creating a negative externality \cite{sefton_information_2001}.

\paragraph{Trustworthy Mechanisms}

To address these problems, scholars have proposed several potential solutions. Drawing on financial market experience, \citeauthor{kyle_continuous_1985} suggests establishing an efficient information processing mechanism to promote healthy data trading market development \cite{kyle_continuous_1985}. \citeauthor{theinternationalconsortiumofinvestigatorsforfairnessintrialdatasharing_fairness_2016} highlights the importance of implementing transparent mechanisms for data sharing and ensuring equitable allocation of access rights \cite{theinternationalconsortiumofinvestigatorsforfairnessintrialdatasharing_fairness_2016}. \citeauthor{li_valid_2019} proposes designing automated structured data exchange mechanisms and developing B2B processes in a regulated environment to increase trustworthiness and decrease costs \cite{li_valid_2019}. \citeauthor{calancea_techniques_2021} further proposes a series of methodologies for effectively managing FAIR-modeled metadata, which are seamlessly integrated into a data marketplace platform to ensure their proper implementation. The absence of a central authority ensures businesses have complete control of the terms and conditions governing the utilization of their data, increasing the system's trustworthiness. 

Blockchain technology demonstrates significant potential in establishing trustworthy mechanisms for data trading. \citeauthor{he_accountable_2019} suggest employing blockchain to establish traceability and evidence preservation mechanisms for credit records \cite{he_accountable_2019}. \citeauthor{oh_deposit_2021} explore decision-making models of data brokers using blockchain, conducting credit analysis while protecting privacy \cite{oh_deposit_2021}. \citeauthor{abubaker_trustful_2022} employs blockchain to establish a distributed, secure, and trustworthy environment for trading big data \cite{abubaker_trustful_2022}.

Additionally, the research integrates secure analysis and performance evaluation methods to assess the viability and efficacy of the trading platform. \citeauthor{he_accountable_2019a} proposes an accountable blockchain-based data trading platform that utilizes secure dataset similarity comparison methods to identify illegal resales before listing datasets on the data market \cite{he_accountable_2019a}. \citeauthor{kosba_hawk_2016} also suggest that it is important to acknowledge that financial transactions cannot be stored in unencrypted form on the blockchain to preserve privacy \cite{kosba_hawk_2016}.

\paragraph{Supply Optimization}

Another significant challenge faced by data element market development is data supply. Valuable data is dispersed among various entities, making it challenging to gather. Lack of unified standards for collecting, cleaning, labeling, and processing data hinders quality assurance. These challenges show that traditional centralized data production methods cannot meet rapid data demands.


Smart contracts have provided a viable solution to address this issue. \citeauthor{xiong_data_2021} propose using smart contracts to codify legal rules and automatically execute contract terms on the blockchain, which can significantly improve the security of data trading \cite{xiong_data_2021}. \citeauthor{tian_optimal_2019} point out that using contract theory to design optimal smart contracts can maximize data seller revenue while ensuring buyer's individual rationality and incentive compatibility \cite{tian_optimal_2019}.  This mechanism is applicable to both complete and incomplete information markets, effectively striking a balance between the benefits of data trading and the costs associated with data privacy. The decentralized characteristic of smart contracts can reduce the reliance on third parties, thereby mitigating the risks associated with fraud and default \cite{luu_making_2016}. However, it is essential to acknowledge that smart contracts are still susceptible to security vulnerabilities. For instance, the occurrence of TheDAO bug resulted in a loss of \$60 million \cite{luu_making_2016}, highlighting the ongoing necessity for enhancing smart contract security.

Crowdsourcing, as a group intelligence production paradigm, also provides new ideas for addressing supply challenges. \citeauthor{huang_efficient_2022} use crowdsourcing to develop data collection and trading platforms \cite{huang_efficient_2022}. The study establishes a data-centric environment for the widespread implementation of AI-enabled Remote Wireless Sensor Networks. \citeauthor{luu_making_2016} further point out that the crowdsourcing model could allow large-scale participation in data collection and labeling, reducing costs and improving quality and diversity \cite{luu_making_2016}. The crowdsourcing model mobilizes social resources to participate in data trading, greatly expanding data sources and application scenarios, which is significant for promoting the development of the big data industry. 


Social sensing provides another promising approach to optimizing data supply. \citeauthor{gu_cbdtf_2024} suggests using social media platforms, smart hardware, and IoT to make the data collection process trustworthy \cite{gu_cbdtf_2024}. The authors present a novel framework for mobile crowdsensing that aims to facilitate distributed and reliable data trading by incorporating a blockchain-based incentive mechanism. \citeauthor{lu_social_2021} highlight the potential of social signals in providing timely, extensive, and rich intelligence about urban dynamics and social behaviors\cite{lu_social_2021}. Social sensing can be leveraged to gather valuable data that might otherwise be difficult to obtain through traditional methods. For instance, \citeauthor{chen_convolutional_2017} demonstrate how social media platforms can be mined for real-time traffic information. This approach aligns well with the crowdsourcing model mentioned earlier, as it taps into the collective intelligence of social media users. Furthermore, \citeauthor{zhang_cyberphysicalsocial_2018} suggests using massive social sensors collected from big data in Cyber-Physical-Social Systems (CPSS) to construct broader data ecosystems. By incorporating social sensing techniques into data trading platforms, market participants can access a wider range of data sources, potentially improving the diversity and real-time nature of available data.


While crowdsourcing promotes data circulation, it also presents challenges such as ensuring fairness and transparency, preventing data abuse or privacy information leakage \cite{liang_survey_2018}, designing effective incentive mechanisms\cite{ramsundar_tokenized_2018a, rasouli_data_2021}, and protecting data copyright \cite{liang_survey_2018}. Future research should focus on optimizing crowdsourcing applications while prioritizing privacy protection and personal rights to maximize data value and social welfare.

\subsubsection{Addressing Compliance Challenge and Costly Transaction}

The lack of institutions and cost problems manifest in outdated data trading architecture and a lack of value transfer mechanisms. Researchers are investigating the potential of incorporating advanced technologies like decentralized currencies and smart contracts into data trading models to address these challenges.

\paragraph{Trading Architecture}

Traditional data management architectures suffer from severe data isolation, excessive coupling of business logic, complex system structures, and limited scalability. Several technical foundations and infrastructures for data trading have been proposed to address these issues.
  
Several technical foundations and infrastructures for data trading have been proposed. \citeauthor{hedenberg_developing_2018} propose developing software tools for data collection, storage, processing, and trading and establishing efficient data exchange platforms \cite{hedenberg_developing_2018}. The study aims to ensure transaction security and data reliability when trading in the blockchain by utilizing a Python module, Scrapy, to automatically collect news and pricing data and then pass it on to a MongoDB database. \citeauthor{sharma_trustworthy_2020} explore an approach to combine data marketplaces and blockchain for fair and independent data marketplaces, where blockchain enables businesses to be decentralized and more secure \cite{sharma_trustworthy_2020}. \citeauthor{niya_itrade_2021} present the design, implementation, and evaluation of "ITrade," a BC-based IoT data trading platform with a highly scalable microservice-based architecture \cite{niya_itrade_2021}.

Technologies such as asymmetric encryption and consensus algorithms are used to increase the efficiency and security of trading architectures. \citeauthor{xiong_data_2021} suggest optimizing consensus mechanisms such as PoS and DPoS to improve currency transaction efficiency and security \cite{xiong_data_2021}. The study presents a solution for enhancing security by implementing a consortium blockchain and smart contracts to achieve audibility, accountability, and integrity in data trading.

\paragraph{Decentralized Payment Systems}
Traditional systems are burdened with numerous challenges in payment and settlement, such as high trading costs, complexities in cross-border payments, and volatile currency values. These limitations render them to globalized and high-frequency trading requirements \cite{delgado-segura_fair_2020,yao_impact_2018}.

Decentralized currencies\cite{ramsundar_tokenized_2018a}, blockchain, smart contracts, and other technologies \cite{lucking_when_2021} are employed to construct data payment systems. \citeauthor{lucking_when_2021} design an open data trading system in vehicular ad hoc networks using distributed ledger technology (DLT), measure the required communication time for data trading between a vehicle and a roadside unit in a real scenario, and estimate the associated cost \cite{lucking_when_2021}. \citeauthor{li_multiblockchain_2023} construct two-layer data trading markets based on data blockchain (DChain) and value blockchain (VChain), and propose pricing mechanisms to balance the payoffs of both suppliers and users. Experiments demonstrate the strength of this research in enhancing data security and trading efficiency \cite{li_multiblockchain_2023}. 

Data currency, as an emerging financial tool, also provides a potential solution. \citeauthor{ramsundar_tokenized_2018a} and \citeauthor{oh_personal_2019} point out that decentralized data markets can be constructed by introducing tokenized data structures. \citeauthor{xiong_smart_2019} and \citeauthor{chenli_fairtrade_2021} further propose mapping data assets to data currencies through smart contracts, realizing the measurability and circulation of data value \cite{xiong_smart_2019,chenli_fairtrade_2021}. \citeauthor{chenli_fairtrade_2021} propose FairTrade, an efficient atomic exchange-based fair exchange protocol for digital data trading, which uses an AES-256 encryption technique to encrypt data and maintain data integrity, and provides incentive to the users and arbitrators to increase user participation and honesty \cite{chenli_fairtrade_2021}. 

\paragraph{Consensus Mechanisms}
The absence of multi-party participation and co-governance mechanisms is also a main issue in the market. The conventional top-down management model struggles to address diverse interests and cannot effectively handle complex market risks.

The consensus mechanism offers a novel approach to address this issue. In the design of consensus mechanisms, it is essential to establish a rational integration of data pricing and data incentive mechanisms. \citeauthor{tian_optimal_2019} propose utilizing blockchain technology to construct a decentralized governance framework that effectively combines efficiency and security\cite{tian_optimal_2019}. This approach establishes a collaborative governance model involving various entities, including government bodies, enterprises, academia, and the general public. Furthermore, it consists in designing appropriate incentive-compatible mechanisms to stimulate active participation.

In summary, cutting-edge technologies such as decentralized currencies and smart contracts provide new possibilities for reducing costs and optimizing trading institutions.

\subsubsection{Addressing Collateral Consequence and Costly Transaction}

A significant paradox in data trading lies in simultaneously realizing data value and safeguarding privacy, intellectual property rights, and other personal rights. In light of this concern, researchers aim to enhance incentive mechanisms and data management by exploring federated learning, encryption, and cloud computing technologies. The objective is to realize the data's value while ensuring privacy protection.

\paragraph{Incentive Mechanisms}

Traditional data trading either requires centralized data storage, which risks privacy leakage, or requires the outflow of original data, which is difficult for data owners to accept. Federated learning provides an innovative solution to this problem by allowing data to be processed locally, only sharing model updates, and protecting data privacy while reducing the cost of data aggregation and analysis. \citeauthor{hu_trading_2020} adopts the game theory to design an effective incentive mechanism in federated learning, which selects users that are most likely to provide reliable data and compensates for their costs of privacy leakage\cite{hu_trading_2020}. They formulate the problem as a two-stage Stackelberg game and solve the game's equilibrium.  

Appropriate federated learning architectures are designed to address various scenarios, such as horizontal and vertical settings\cite{wang_federated_2021,wang_federated_2021a,ouyang_artificial_2023}. \citeauthor{zheng_flmarket_2022} introduces FL-Market, a model marketplace that ensures privacy protection for model buyers and against an untrusted broker. Additionally, the research proposes a deep learning-based auction mechanism within FL-Market to determine the perturbation levels of local gradients intelligently. Furthermore, an optimal aggregation mechanism is developed for aggregating these perturbed gradients.
\citeauthor{li_novel_2022} further propose using blockchain to construct incentive mechanisms and trust mechanisms within the federation, thereby fostering active participation \cite{li_novel_2022}. 
\citeauthor{khan_federated_2020} use Stackelberg game modeling to design incentive mechanisms in federated learning \cite{khan_federated_2020}.

At the data level, federated feature engineering and federated inference can also be devised to facilitate the circulation and fusion of data \cite{li_survey_2021}. Smart contracts and federated learning contribute significantly to enhancing the security and efficiency of data trading. Smart contracts bolster security by automating contract term execution and safeguarding transaction privacy. Federated learning improves efficiency by distributing model training and preserving data privacy, particularly when handling substantial volumes of sensitive data. However, \citeauthor{mathews_federated_2022} point out that federated learning faces challenges such as high communication costs and adversarial attacks \cite{mathews_federated_2022}. The research offers a methodical synopsis and intricate classification of federated learning, examining the security hurdles while presenting an all-encompassing overview of established defense techniques against data poisoning, inference attacks, and model poisoning attacks. Combining the automation functionalities of smart contracts with the data privacy protection capabilities offered by federated learning can yield a more robust and streamlined solution for data trading.

\paragraph{Data Management}

Novel computing architectures of data management like cloud computing offer significant potential for safeguarding data privacy and unlocking its value. \citeauthor{yue_jcdta_2018} propose the JointCloud Computing Data Trading Architecture (JCDTA) to ensure the security and reliability of data trading in the JointCloud environment and maintain the value invariance of data resources \cite{yue_jcdta_2018}. The study uses blockchain decentralization and openness to establish a big data transaction platform for multiple users, guarantees the security and reliability of data trading in the JointCloud environment, and ensures the value invariance of data resources.

Metadata management also plays a pivotal role in striking a balance between data value and privacy. Metadata management constructs an "ID" for data resources, bringing massive data assets into a refined management track. It is imperative to devise an integrated metadata framework that caters to diverse data types, including structured, semi-structured, and unstructured data. \citeauthor{hayashi_understanding_2020} point out that identifying, classifying, retrieving, and tracing data based on metadata will significantly improve the precision and efficiency of data asset management, providing basic support for data trading, pricing, metering, auditing, and other aspects \cite{hayashi_understanding_2020}. \citeauthor{lawrenz_significant_2020} further outline the significance of metadata in data trading, and also discuss the role of metadata in terms of data quality \cite{lawrenz_significant_2020}.

Constructing a data management system consisting of data collection, storage, retrieval, analysis, and application is crucial to alleviate the consequences and cost. \citeauthor{zhou_securing_2023} proposes building a complete lifecycle data management and promoting efficient data trading \cite{zhou_securing_2023}. The study introduces a trading model for big data based on Ethereum, which aims to establish a comprehensive and secure system. The objective is to provide users with more convenient, secure, and professional services. Additionally, the study proposes the inclusion of a trusted third-party entity that offers professional data evaluation services. This entity actively safeguards users' data ownership in case of disputes.

In summary, the evolution of research in data trading has progressed from single-point methods to integrated solutions, addressing compliance challenges, collateral consequences, and costly transactions. The literature review framework, derived from the evolving research paradigm, explores technologies in institutional construction, collateral consequences control, and cost management. Furthermore, cutting-edge research proposes solutions for jointly addressing multiple problems, such as constructing trust mechanisms, optimizing data supply, reshaping trading architecture, building decentralized payment systems, improving incentive mechanisms, and optimizing data management. These efforts collectively contribute to developing a mature and efficient data trading market.

\section{Research Trends Analysis} \label{casestudy}

\subsection{Research Trends}

We have identified a research trend shift from single-pointed methods to integrated solutions through a systematic review and analysis of existing data trading research. Research in the early stage mainly focused on one problem, such as enhancing laws and regulations to compensate for compliance challenges, adopting privacy-preserving computing technology to control collateral consequences, and reducing transaction costs through data pricing. These studies have made certain breakthroughs in their respective fields, enhancing people's comprehension of data trading. However, the current state of data trading reveals that isolated technological advancements are insufficient in addressing the problems associated with data trading. Due to the inherent complexity of data trading, which often involves multiple stakeholders, problems related to compliance, consequences, and costs become intertwined. Consequently, it poses a significant challenge for any individual technology to resolve intricate systemic problems effectively.

As researchers delve further into the complexities of data trading, they increasingly acknowledge that 3C problems are not isolated concerns but interconnected. The compliance challenge exacerbates collateral consequences, while collateral consequences problems amplify transaction costs; consequently, high transaction costs impede the improvement of the compliance system.

\begin{figure}[htbp]
    \centering
    \includegraphics[width=0.7\linewidth]{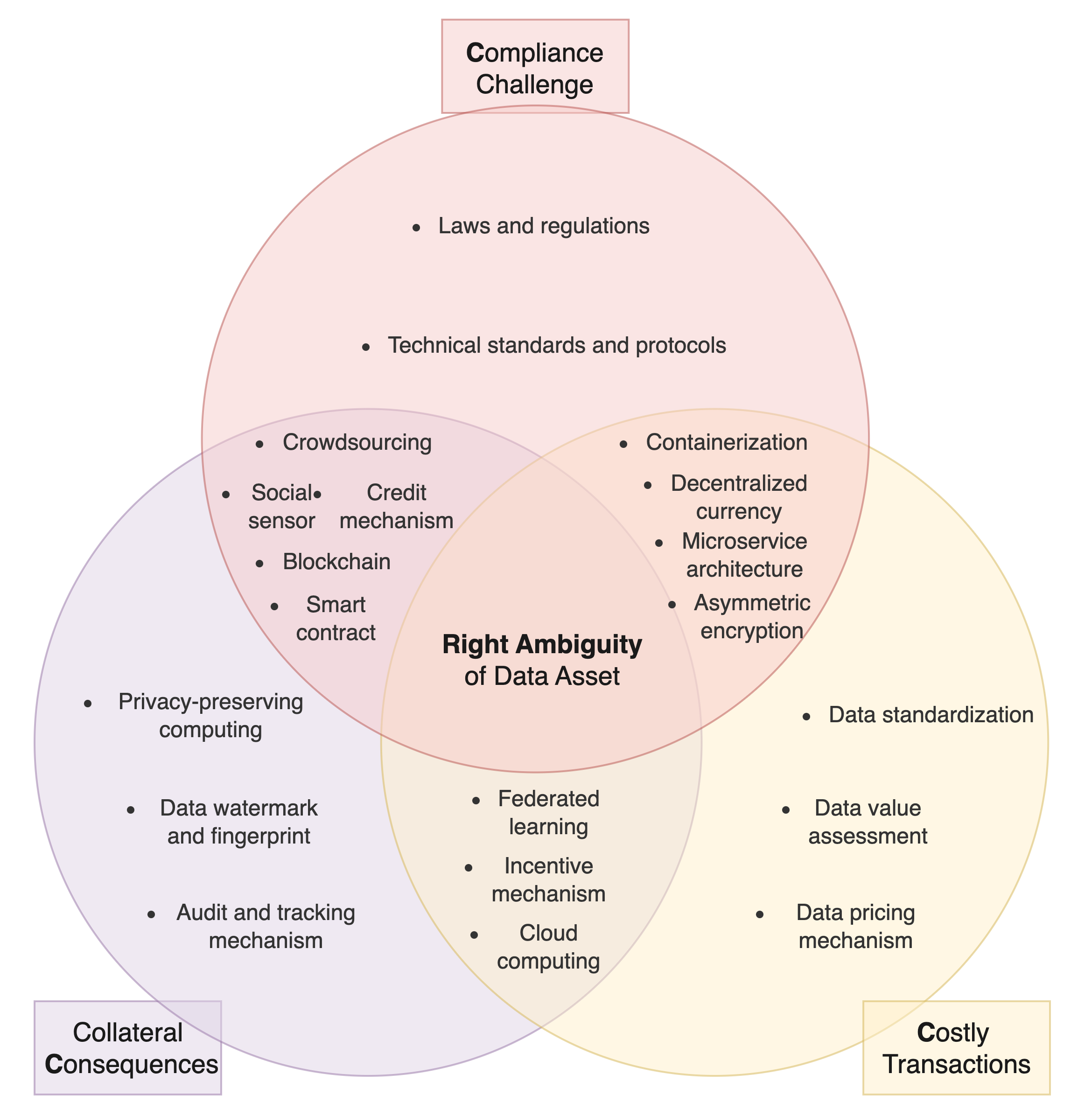}
    \caption{3C problems and corresponding technologies}
    \label{fig:all_methods}
\end{figure}

The correlation of 3C problems drives the evolution of research patterns, necessitating the utilization of integrated solutions and multidimensional collaborative technologies. For example, blockchain and privacy-preserving computing technologies can be integrated to promote privacy protection and institutional construction jointly, decentralized trading mechanisms can optimize compliance and reduce transaction costs, and federated learning and deep learning can be employed to maximize data value while protecting privacy. We systematically review and analyze related research methods as shown in Fig. \ref{fig:all_methods}, and find out that data trading research is moving from single-point technological innovation to integrated system solutions. This transformation corresponds to the complexities of data trading, highlighting the increasing refinement of research in this field.

\subsection{Research Gaps}

\begin{figure}[htbp]
    \centering
    \includegraphics[width=1\linewidth]{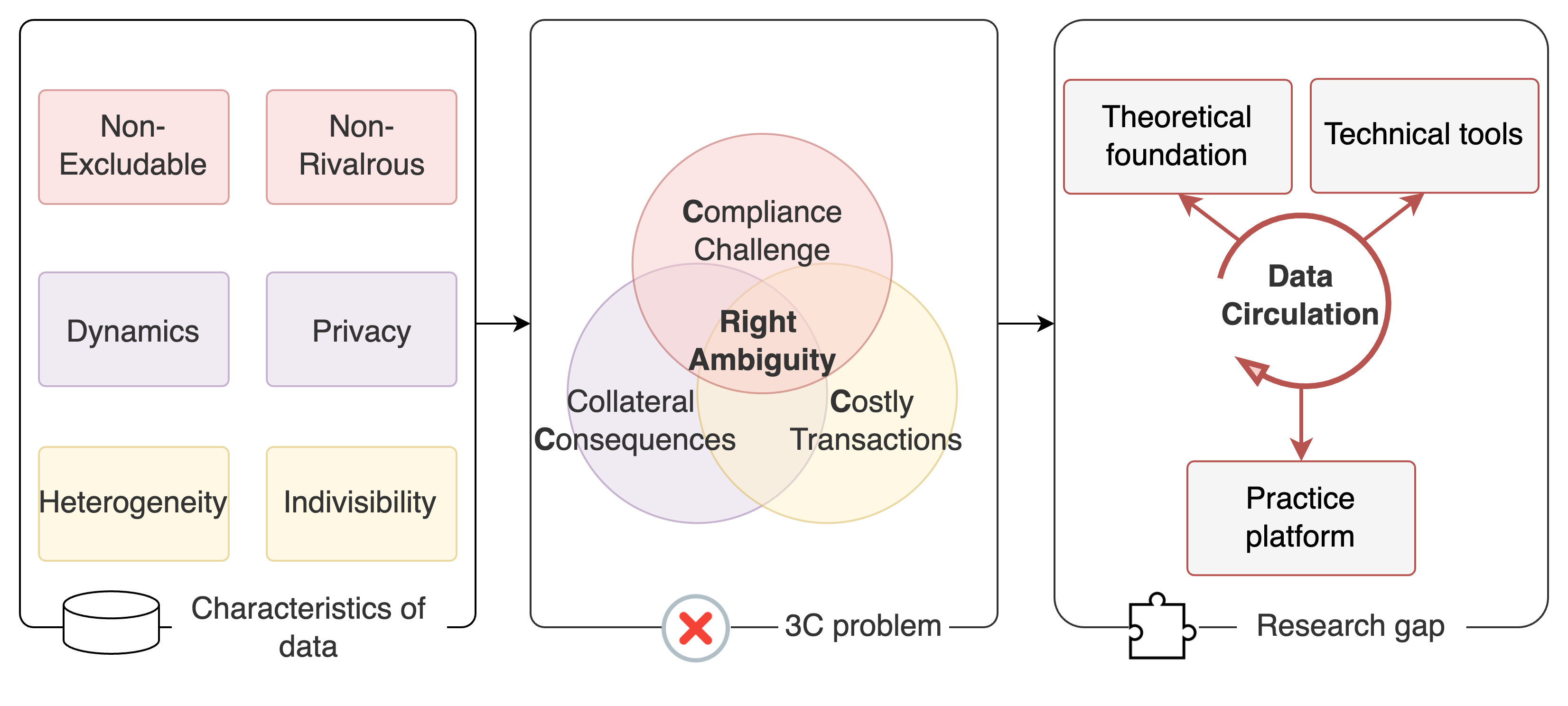}
    \caption{Analysis of data characteristics causing fundamental obstacle of right ambiguity and 3C problems, highlighting research gaps in data circulation}
    \label{fig:data_character}
\end{figure}

Although the expansion of research perspectives and innovative research methods have paved the way for addressing data trading problems,  it is crucial to recognize that existing studies still exhibit certain constraints:

First, the theoretical foundation is not sufficiently solid. The mismatch between data and theories in economics, sociology, and other fields leads to the inability of existing theories to guide data trading effectively. As an emerging factor of production, data has characteristics that are significantly different from traditional factors, such as non-exclusivity, sensitivity, replicability, indivisibility, and heterogeneity. These characteristics render data susceptible to Right Ambiguity, making applying conventional factor market theories in data trading challenging. This mismatch gives rise to the 3C problems depicted in Fig. \ref{fig:data_character}, accompanied by various associated phenomena.

Second, the systematicness of technical tools remains inadequate. Each step in data circulation, including data collection, storage, analysis, pricing, trading, security, and privacy protection, presents complex technical challenges. Addressing these challenges necessitates collaborative innovation across various domains, such as cryptography, federated learning, deep learning, and blockchain technologies. However, we note that many studies are still limited to exploring a specific technical field. Furthermore, there is inconsistency in standards and inadequate connectivity within different technical modules. As a result, individual technological advancements face challenges when attempting to fulfill their intended purpose within the extensive data trading system. The absence of a comprehensive perspective on the tool system hinders breakthroughs in data trading.

Furthermore, the establishment of data trading platforms remains significantly delayed. These platforms are essential intermediaries connecting data's supply and demand, which is crucial in unlocking data value. However, due to the absence of unified industry standards and regulatory norms, the development of trading platforms is scattered and fragmented. It is imperative to enhance mutual trust mechanisms and interoperability to address the "data solios" issue between different platforms. This highlights the necessity for well-established full-lifecycle data trading platforms that are mature, standardized, and trusted.

The root of the above research limitations lies in the fact that data, as an emerging factor of production, still needs further exploration and research. The theory, technology, and practice have not adequately adapted to the profound influence of the substantial variable of "data" on the economy and society. This also highlights another constraint of existing research: the incorporation of interdisciplinary and cross-field studies remains in its nascent stage. It is crucial to transcend conventional boundaries by embracing a more inclusive and open perspective to promote interdisciplinary crossover and expand our understanding. This necessitates restructuring the framework of data trading with a systematic approach.

These research limitations and the 3C problems in data trading can all be attributed to the data's Right Ambiguity. To address this ambiguity, future research needs to systematically understand data as a factor of production from a new perspective.

\subsection{Data Usufruct Research Framework}

\subsubsection{Data Usufruct}

Data usufruct, which refers to the entitlement of individuals who do not own data to utilize and benefit from data \cite{shen2020data}, provides a new theoretical perspective for data trading. With the increasing importance of data as a new factor of production, traditional ownership-based legal protection becomes challenging to apply. Conventional research is based on the exchange of data ownership where individuals or organizations acquire full disposal rights over data assets. However, due to the non-rivalrous and non-excludable data characteristics, trading ownership encounters various shortcomings such as vulnerability to copying and reselling for arbitrage purposes. These limitations impede the development of a robust market. To address this issue, \citeauthor{shen2020data} proposed the concept of "data usufruct" by separating ownership and usufruct of data \cite{shen2020data}. 
Similar to commodity usufruct, data usufruct endows data with certain attributes, enabling control over information leakage during data trading while effectively addressing copyability and resellability issues.

However, the implementation of data usufruct faces several practical challenges. These challenges encompass difficulties in measurement and calculation, prevention of misuse, low efficiency in circulation, unclear attribution of rights, inadequate legal framework, and insufficient technical support. To address these challenges, it is imperative to propose a computable framework for data usufruct. Additionally, standardized data usufruct units should be constructed based on privacy-preserving computing technology. Through privacy-preserving computing technologies like differential privacy, it becomes feasible to quantify the value generated from data usufruct. This enables a pricing and cost assessment akin to that of conventional commodities, ensuring the market's sustainable growth. Furthermore, usufruct attribution should be established following the non-arbitrage principle. Ultimately, separating data ownership and usufruct will lead to the development of a dynamic pricing mechanism and stimulating data circulation.


Based on data usufruct and existing research, the following section proposes the future research paradigm of data trading and provides definitions and quantification methods for 3C problems \cite{yu_swdpm_2023}.

\subsubsection{Problem Definition}

Trading behavior in the data market can be modeled as Integrated Markov Decision Processes ($\mathcal{IMDP}$) containing Hidden Markov Models (HMM). Each trader participating in the transaction (such as buyers and sellers) can be regarded as an independent Markov Decision Process (MDP), and their interactions and influences constitute the integration of the entire system.

For each trader, actions \( a_t^i \) are made based on its current state \( s_t^i \) and observation \( o_t^i \). These decisions may include purchasing, selling, or other trading behaviors. Meanwhile, transactions and competition between subjects also affect their state transitions \( P(s'|s,a) \) and reward \( R^i(s,a) \), thus forming a complex integrated system.

For all traders $\mathcal{T}={1,2,...,N}$ (with $N$ traders), the Integrated Markov Decision Process can be represented as $\mathcal{IMDP} = (\mathcal{S}, \mathcal{A}, \mathcal{P}, \mathcal{R}, \mathcal{O}, \Gamma, \Pi)$, where:

\begin{itemize}
\item $\mathcal{S}$ is the state space set. For trader $i$, its state $s_t^i$ at time step $t$ can be represented as a vector $[v_t^i, c_t^i]$, where $v_{t}^{i}$ represents the data state owned by trader $i$ at time step $t$; $c_{t}^{i} \in \mathbb{R}_{\geq 0}$ represents the amount of money owned by trader $i$ at time step $t$. From a classification perspective, states include Sold, Hold, and Bought.
\item $\mathcal{A}$ is the action space set. For trader $i$, its action $a_t^i$ at time step $t$ can be represented as a vector $[dv_t^i, dc_t^i]$, where $dv_t^i$ represents the operation performed by trader $i$ on the data at time step $t$; $dc_t^i \in [-c_t^i, \infty)$ represents the change in the money traded by trader $i$ at time step $t$.
\item $\mathcal{P}$ represents the state transition probability. Given the set of trader states $\mathbf{S}_t = \{s_t^1, s_t^2, \ldots, s_t^N\}$ and the set of actions $\mathbf{A}_t = \{a_t^1, a_t^2, \ldots, a_t^N\}$, the probability of trader $i$ transitioning from $s_t^i$ to $s_{t+1}^i$ is $P(s_{t+1}^i|s_t^i,a_t^i)$. The state transition probability of the entire system is:
\begin{equation}
\mathcal{P}(\mathbf{S}_{t+1}|\mathbf{S}_t, \mathbf{A}_t) = \text{Product}\left(P(s_{t+1}^i | s_{t}^i, a_t^i), \text{ for } i \in \mathcal{T}\right)
\label{eq:state_transition}
\end{equation}
\item $\mathcal{R}$ is the reward function, representing the immediate reward obtained after taking each action in each state.
\item $\mathcal{O}$ is the observation space set, representing the environmental information each trader can observe, including Bull, Stable, and Bear.
\item $\Gamma$ is the discount factor, representing the discount rate of future rewards.
\item $\Pi$ is the policy set, representing the action strategy taken by each trader in each state.
\end{itemize}

\subsubsection{Manifestation of the 3C Problems}

\begin{figure}[htbp]
    \centering
    \includegraphics[width=1\linewidth]{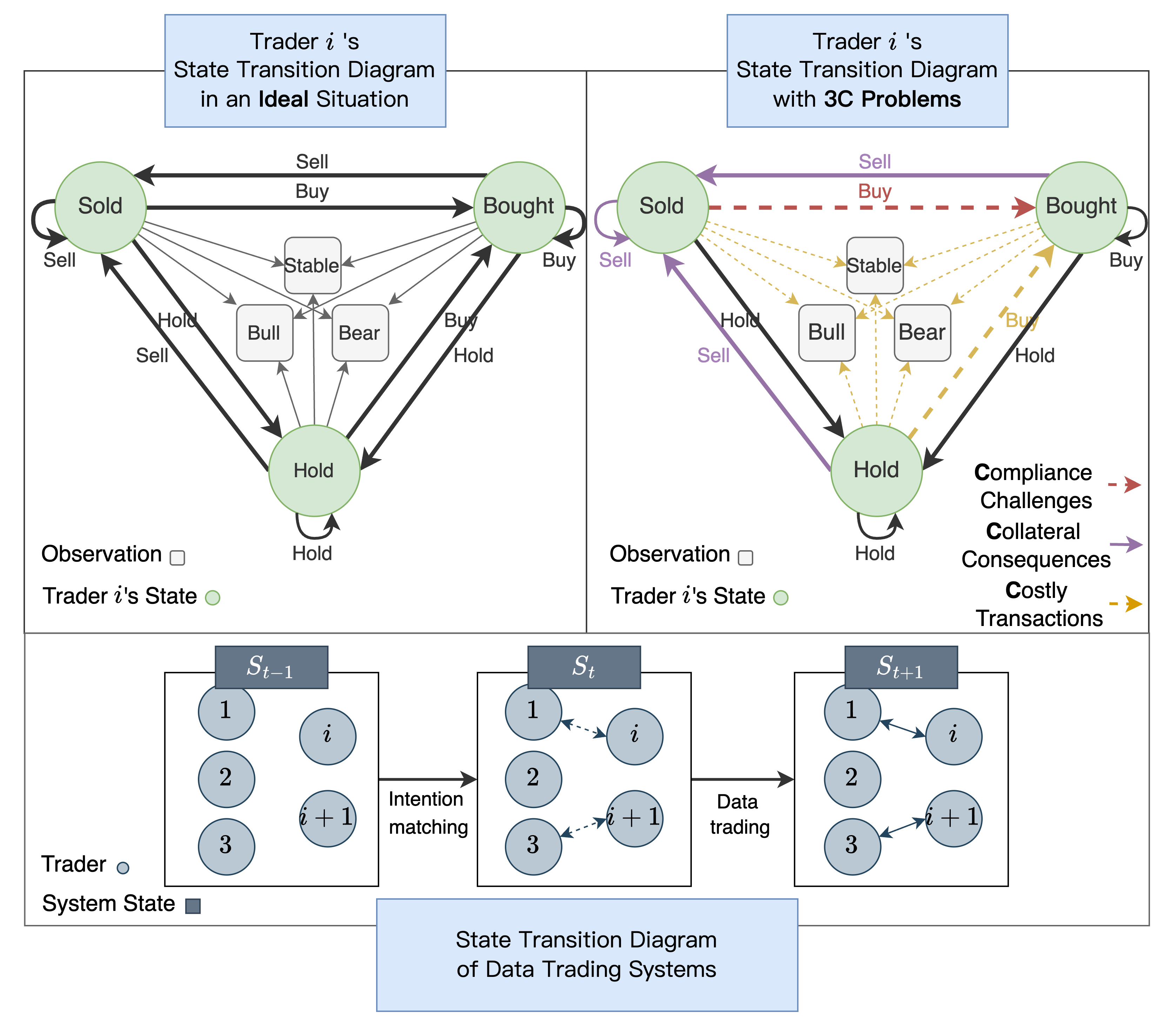}
    \caption{State transition diagram of the data trading system and trader \(i\), illustrating regular and 3C problems-related state transitions: green circles represent the states of traders, including Sold, Hold, and Bought; gray rectangles represent the traders' observations of the market, including Bull, Stable, and Bear. The transitions between states are represented by black solid arrows, with the trading behavior (such as buying or selling) involved in each state transition labeled.}
    \label{fig:definition}
\end{figure}

Based on the above modeling, the state transition diagram of data trading and the manifestation of 3C problems are shown in Fig. \ref{fig:definition}.

The 3C problems in MDP include are presented as:
\begin{itemize}
\item Compliance Challenges as limited state transitions: Due to the lack of a system, the state transition probability $\mathcal{P}(s_{t+1} | s_t, a_t)$ is restricted, and traders cannot accurately expect the results of their behaviors, leading to uncertainty and instability of state transition probabilities. For example, the lack of data privacy protection regulations may make traders reluctant to participate in transactions, thus limiting market liquidity and vitality. The Compliance Challenges are represented by the red dashed line in Fig. \ref{fig:definition}.
\item Collateral Consequences as imbalanced reward function: Negative externalities cause an imbalance in the reward function $\mathcal{R}(s_t, a_t)$, causing traders to fail to fully consider long-term social benefits and hindering the market's healthy development. For example, a trader illegally selling data may gain short-term benefits, but damage the overall trust and sustainability of the data market. The Collateral Consequences are shown by the purple dashed line in Fig. \ref{fig:definition}.
\item Costly Transactions as limited action and observation space: High transaction costs limit the action space $\mathcal{A}$ and observation space $\mathcal{O}$ of traders, making it difficult for traders to conduct effective transactions and obtain information, thus limiting market liquidity and transparency. For example, high data purchase costs may make it difficult for traders to conduct frequent transactions. The Costly Transactions are shown by the orange dashed line in Fig. \ref{fig:definition}.
\end{itemize}

\subsubsection{Quantification of the 3C Problems}

To better quantify 3C problems, we propose an assessment method to evaluate their impact on the data trading market.

\paragraph{Compliance Challenges Metrics}

The Compliance Challenges arose from the lack of regulations and standards, causing high trading barriers, and limiting traders' state transitions. Therefore, the compliance challenges can be measured by quantifying the imbalance of state transition probabilities. The impact of the lack of a system on state transitions can be reflected by calculating the entropy of each state's transition probability distribution. The degree of imbalance of state transitions \( \Phi_P \) is defined as:

\begin{equation}
\Phi_P = \sum_{i \in \mathcal{T}} \sum_{s \in \mathcal{S}} \mathcal{H}(\mathcal{P}(s_{t+1}^i|s_t^i, a_t^i))
\label{eq:phi_p}
\end{equation}
where:
\begin{equation}
\mathcal{H}(\mathcal{P}(s_{t+1}^i|s_t^i, a_t^i)) = - \sum_{s_{t+1}^i \in \mathcal{S}} \mathcal{P}(s_{t+1}^i|s_t^i, a_t^i) \log \mathcal{P}(s_{t+1}^i|s_t^i, a_t^i)
\label{eq:entropy}
\end{equation}

The higher the entropy, the more uniform the distribution of state transitions, the smaller the impact of the compliance challenges may be, and vice versa.

\paragraph{Collateral Consequences Metrics}

Collateral Consequences manifested in the widespread social concern and controversy of data trading since the external effects are not sufficiently compensated, causing an imbalance in the reward function \( \mathcal{R}^i(s^i, a^i) \). We can quantify the degree of imbalance of the reward function \( \Phi_R \) as:
\begin{equation}
\Phi_R = \sum_{i \in \mathcal{T}} \sum_{s^i \in \mathcal{S}} \sum_{a^i \in \mathcal{A}} \left| \mathcal{R}^i(s^i, a^i) - \mathcal{R}^i_{\text{optimal}}(s^i, a^i) \right|
\label{eq:phi_r}
\end{equation}
where \( \mathcal{R}^i(s^i, a^i) \) is the actual immediate reward for trader \(i\). \( \mathcal{R}^i_{\text{optimal}}(s^i, a^i) \) is the socially optimal reward for trader \(i\) considering collateral consequences.

\paragraph{Costly Transactions Metrics}

Costly transactions manifest in various costs in negotiation, contracting, and enforcement, limiting the action space \( \mathcal{A} \) and observation space \( \mathcal{O} \)of traders. The degree of limitation of action and observation space \( \Phi_S \) is defined as:
\begin{equation}
\Phi_S = \Phi_{\mathcal{A}} + \Phi_{\mathcal{O}}
\label{eq:phi_s}
\end{equation}
where:
\begin{equation}
\Phi_{\mathcal{A}} = \sum_{i \in \mathcal{T}} \sum_{s^i \in \mathcal{S}} \left( 1 - \frac{|\mathcal{A}^i(s^i)|}{|\mathcal{A}^i_{\text{max}}(s^i)|} \right)
\label{eq:phi_a}
\end{equation}
\( |\mathcal{A}^i(s^i)| \) represents the number of available actions for trader \( i \) in state \( s^i \), and \( |\mathcal{A}^i_{\text{max}}(s^i)| \) represents the maximum number of available actions for trader \( i \) in state \( s^i \) without cost limitations.

\begin{equation}
    \Phi_{\mathcal{O}} = \sum_{i \in \mathcal{T}} \sum_{s^i \in \mathcal{S}} \left( 1 - \frac{|\mathcal{O}^i(s^i)|}{|\mathcal{O}^i_{\text{max}}(s^i)|} \right) 
\end{equation}
\( |\mathcal{O}^i(s^i)| \) represents the number of available observations for trader \( i \) in state \( s^i \), and \( |\mathcal{O}^i_{\text{max}}(s^i)| \) represents the maximum number of available observations for trader \( i \) in state \( s^i \) without cost limitations.

\paragraph{Comprehensive Metrics} 
Combining the above three quantitative indicators, the overall severity of the 3C problems \( \Phi \) is obtained as:
\begin{equation}
    \Phi = \alpha \Phi_P + \beta \Phi_R + \gamma \Phi_S \label{eq:phi}
\end{equation}
where \( \alpha \), \( \beta \), and \( \gamma \)  are weight coefficients used to adjust the influence of each problem on the overall data trading system.

Defining these metrics enables a precise measurement of data trading systems, providing a robust framework for evaluating performance. The Compliance Challenges Metric (\(\Phi_P\)) assesses the impact of regulations, identifying areas for regulatory enhancement. The Collateral Consequences Metric (\(\Phi_R\)) quantifies social and economic imbalances, guiding the establishment of socially optimal reward structures. The Costly Transactions Metric (\(\Phi_S\)) highlights inefficiencies in action and observation spaces. Collectively, these metrics offer a toolset for identifying weaknesses, tracking progress, and directing future optimization in data trading systems.

\subsection{Future Directions}

Data, as an emerging factor of production, derives its value from unrestricted circulation and effective utilization. However, this notion inherently contradicts the concept of private ownership. Like the capitalist system of private ownership, the private ownership of data can impede data circulation and restrict the maximization of its value. Therefore, future research needs to focus on resolving the contradiction between data private ownership and data circulation, exploring new data rights models, and systematically addressing the 3C problems to promote the efficient circulation and trading of data.

The introduction of data usufruct, as suggested earlier, holds the potential to address this contradiction effectively. Data usufruct aims to transform data trading from traditional ownership to the usufruct. Non-data owners could use data within an agreed period by paying a certain fee, without the need to purchase data ownership, which simplifies the data acquisition and trading process, reduces data leakage or abuse risks, and promotes the circulation of data. Simultaneously, data usufruct also offers a novel theoretical foundation for developing data pricing and trading mechanisms, thereby facilitating the establishment of an organized and transparent data market.

\begin{figure}[htbp]
    \centering
    \includegraphics[width=1\linewidth]{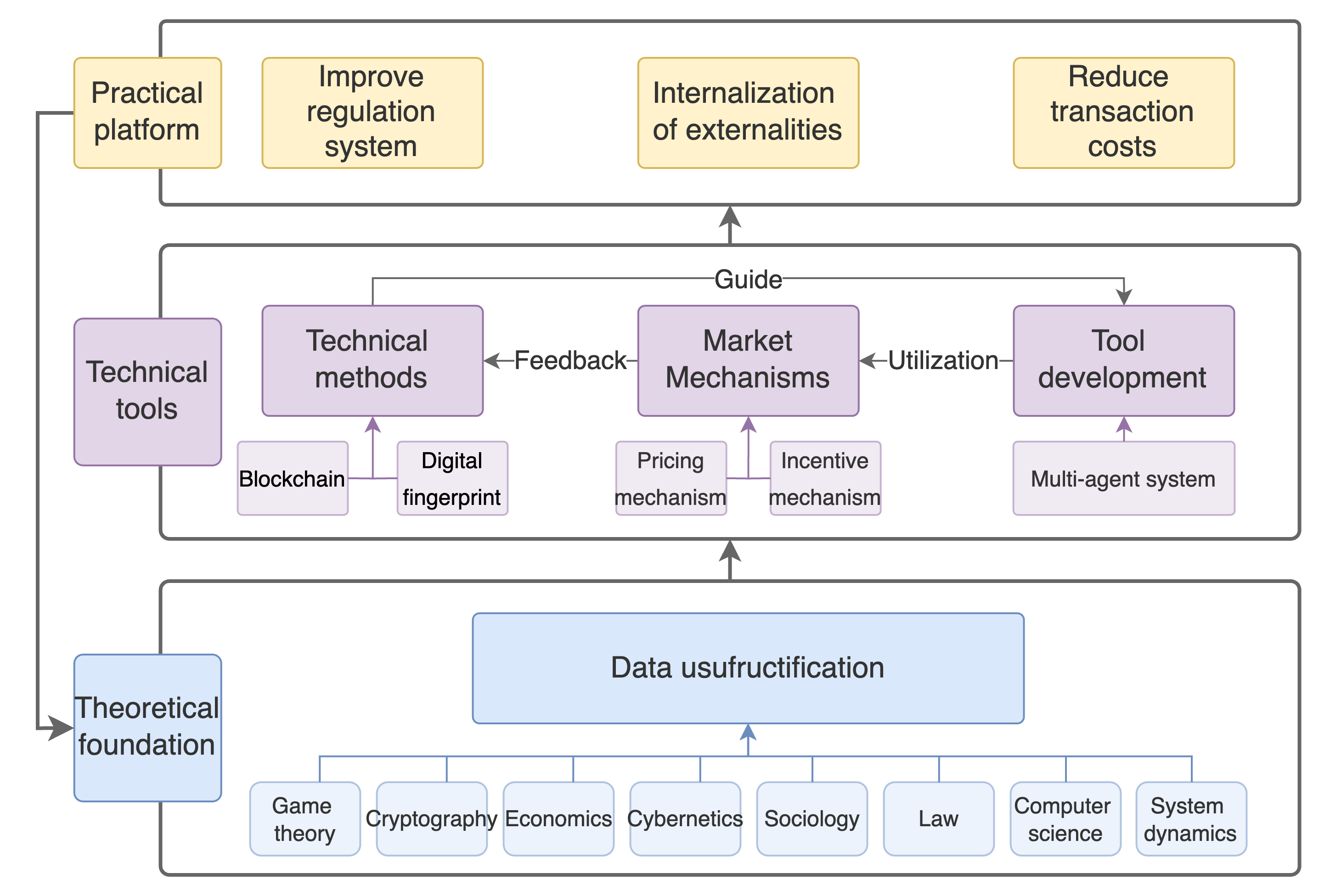}
    \caption{Research framework of data usufruct}
    \label{fig:framework}
\end{figure}

Under this research paradigm, future research needs to explore the theoretical foundation, technical tools, and practical platforms of data usufruct, as shown in Fig. \ref{fig:framework}. At the theoretical level, it is urgent to clarify the legal, economic, and technical attributes of data usufruct, depict the allocation rules and circulation mechanisms, and construct a systematically complete theoretical framework. This requires the interdisciplinary integration of law, economics, data science, and other disciplines to break barriers and provide theoretical support. Clarifying the legal aspects involves establishing clear definitions and protections for data usufruct, ensuring that the rights and responsibilities are well defined and enforceable. Economically, it is important to usufruct attribution models that capture characteristics and align with non-arbitrage principles. Understanding the mechanisms that can support trading and ensuring they are secure and efficient is critical. This interdisciplinary approach will help create a robust theoretical foundation that addresses the multifaceted nature of data usufruct.

At the technical level, exploring the implementation approaches for data usufruct is necessary. A mechanism for confirming and authorizing data usufruct can be developed utilizing privacy-preserving computing, watermarking, blockchain, and other technologies. Privacy-preserving computing techniques, such as differential privacy, will be essential to make usufruct computable. Watermarking can help protect the integrity and origin of data usufruct, making it difficult for unauthorized parties to misuse it. Blockchain technology can provide a transparent and immutable record of data usufruct transactions, ensuring trust and security in the trading process. Additionally, employing smart contracts and other technologies can enable automated trading and secure circulation. Integrating multiple innovative technologies will provide a robust technical foundation for trading data usufruct.

At the practical level, constructing a data usufruct trading platform will become an important way to apply theory and technology. The objective is to establish mechanisms supporting data usufruct registration, trading, settlement, and dispute arbitration. This will facilitate the creation of an efficient, transparent, and trustworthy ecosystem for data usufruct trading. The ultimate goal is to promote the practical application of this innovative theory. Platform construction should leverage the expertise of established data trading platforms, enhance trading rules and regulatory measures, and establish a well-organized and equitable trading environment. Furthermore, the data usufruct trading platform will function as an experimental domain for theoretical and technological advancements, fostering the ongoing enhancement of theory and the iterative progression of technology.

\section{Conclusion and Future Works} \label{conclusion}

In this study, we summarize the problems in data trading as 3C problems, conduct a comprehensive review and assessment of data trading research, and propose a data usufruct research paradigm as a potential solution to the 3C problems. 

Due to the special characteristics of data, such as non-exclusivity, non-rivalry, dynamics, and heterogeneity, data trading faces many problems, including Compliance Challenges, Collateral Consequences, and Costly Transactions, which we call 3C problems. These problems are intertwined, forming a complex systemic problem. Our analysis reveals a shift in research paradigm from individual approaches towards integrated solutions: In the early stages, methods like revising laws and regulations and formulating industry standards are proposed to improve institutional construction; Privacy computing, data watermarking, and other technologies are used to control collateral consequences; Data pricing models and data standardization are promoted to reduce transaction costs. More recently, integrated solutions have been suggested, including establishing trustworthy mechanisms, reshaping transaction architectures, and constructing decentralized platforms for data trading. Despite various attempts, effectively tackling the 3C problems remains challenging due to their complexity. Through our review, we identified a lack of comprehensive understanding regarding the intrinsic characteristics of data in existing research, which consequently hinders an accurate apprehension of the fundamental obstacles in data trading - Right Ambiguity. Right Ambiguity refers to unclear definitions of data's property rights, entitlements, interests, and responsibilities, which leads to 3C problems and further hampers data circulation. 

Data usufruct represents a potential solution to address the fundamental obstacle in data trading. Data usufruct refers to individuals who do not own data to utilize and benefit from it.
Through data usufruct, it is expected that data can share attributes similar to conventional commodities, transforming data into a mature commodity. Consequently, it becomes feasible to establish a comprehensive data usufruct trading framework by leveraging well-established economic theoretical systems. For example, data usufruct can be traded like financial options, where data buyers obtain the right to use data within an agreed period by paying a usufruct fee rather than purchasing ownership. This concept can simplify the data trading process and reduce the risks that may arise from data leakage or abuse.

The unrestricted data circulation through data usufruct holds the potential to address 3C problems. Moreover, it can significantly influence various industries downstream by facilitating the emergence of innovative applications and business models:

\begin{itemize}
\item Multi-source data fusion and innovation: The circulation of data facilitates the fusion of multi-source data in various fields such as transportation, finance, and healthcare. For instance, intelligent optimization and accurate traffic flow prediction can be achieved within the transportation domain by integrating real-time data from road sensors, vehicles, and smartphones. This integration significantly enhances traffic management efficiency and improves travel experiences. Similarly, in the medical field, integrating multi-source data from hospitals, patients, and wearable devices can drive advancements in personalized medicine and precision medicine while enhancing the quality and effectiveness of medical services.
\item Intelligent manufacturing and industrial Internet: Data circulation promotes the interconnection of information between equipment, systems, and enterprises, realizing the collaborative optimization of the entire industry chain. For example, by sharing real-time data, manufacturing enterprises can achieve intelligent monitoring and predictive maintenance of production equipment. Consequently, this reduces equipment failure rates and downtime while enhancing production efficiency and product quality. Simultaneously, data circulation enhances supply chain transparency and optimization, thereby improving the efficiency and flexibility of supply chain management.
\item AI security and management: The circulation of data presents novel opportunities for managing AI systems and guiding human behavior. By incorporating multi-source data, AI systems can make more precise decisions and predictions, thereby enhancing the security and reliability of these systems. Furthermore, data circulation facilitates transparency and traceability in AI technology, contributing to establishing a transparent management system for AI. This fosters trustworthy AI while also promoting its widespread application.
\item Next-generation AI facilities and multi-agent systems: Data circulation provides opportunities for AI facilities and multi-agent systems, enabling collaborative work among agents and enhancing system intelligence and response speed. For example, in smart city management, multi-agent systems can work collaboratively based on real-time data to improve urban operation efficiency and residents' quality of life. In addition, data circulation helps to build an open AI ecosystem and promote innovation and technological progress.
\end{itemize}

To facilitate data circulation, it is essential to carry out research on data usufruct in future work, as shown in Fig. \ref{fig:framework}. Theoretically, a comprehensive foundation should be established by integrating multiple disciplines. Technically, developing tools such as data watermarking for embedding usufruct information, blockchain for registration, confirmation, and authorization, and employing smart contracts and privacy-preserving computing for secure and automated circulation of usufruct is necessary. Practically, a trading platform should be built where the market mechanism determines prices and trading occurs transparently, addressing existing 3C problems. This integrated approach of theory, technology, and platform construction based on data usufruct promises a comprehensive resolution of 3C problems, fostering a healthy data market.

\newpage
\balance

\printbibliography

\newpage
\onecolumn

\appendix

\section{Supplementary Tables}

This appendix provides supplementary tables summarizing representative publications addressing the 3C problems in data trading. Tables \ref{tab:system_construction}, \ref{tab:negative_externality_control}, \ref{tab:cost_management}, and \ref{tab:method_summary} offer a comprehensive overview of the strategies, methods, and key literature discussed in the main manuscript.

\begin{table*}[h!]
\centering
\caption{Institutional construction for data trading}
\label{tab:system_construction}
\begin{tabular*}{\linewidth}{@{\extracolsep{\fill}}p{3cm}p{3cm}p{6cm}p{3cm}}
\toprule
Problem & Existing Strategies & Implementation Methods & Representative Literature \\ \midrule
\multirow{4}{2.5cm}{\parbox{2.5cm}{Unclear data property rights}} & Revise and improve laws and regulations & Data rights registration system & \cite{gao_data_2020}, \cite{gong_research_2019}, \cite{liu_docs_2022} \\ \cmidrule{3-4}
& \multirow{3}{*}{} & Hierarchical authorization based on importance and sensitivity & \cite{xu_bhda_2020}, \cite{xuan_hierarchically_2019} \\ \cmidrule{3-4}
& & Distinguish data ownership, usage rights, and disposal rights & \cite{_china_} \\ \cmidrule{3-4}
& & International cooperation and co-governance & \cite{shabani_data_2021}, \cite{burri_data_2021}, \cite{burri_governance_2017}, \cite{mishra_building_2018} \\ \midrule
\multirow{4}{2.5cm}{\parbox{2.5cm}{Lack of data usage norms}} & Formulate technical norms and industry standards & Standards for data encryption, anonymization, interfaces, etc. & \cite{arenas_data_2013}, \cite{fagin_data_2003}, \cite{suzuki_datatrading_2021} \\ \cmidrule{3-4}
& & Formulate norms for cross-border data circulation & \cite{burri_data_2021} \\ \bottomrule
\end{tabular*}
\end{table*}

\begin{table*}[h!]
\centering
\caption{Controlling Collateral Consequences of data circulation}
\label{tab:negative_externality_control}
\begin{tabular*}{\linewidth}{@{\extracolsep{\fill}}p{3cm}p{3cm}p{6cm}p{3cm}}
\toprule
Problem & Existing Strategies & Implementation Methods & Representative Literature \\ \midrule
\multirow{3}{3.5cm}{\parbox{3.5cm}{Contradiction between value release and privacy protection }} & Privacy computing & Applying privacy-preserving computing techniques in data collection, processing, analysis, and trading & \cite{xiong_effective_2020}, \cite{kellaris_practical_2013}, \cite{xue_blockchainbased_2023}, \cite{zhao_machine_2019}, \cite{gao_privacypreserving_2020}, \cite{garrido_revealing_2022} \\ \addlinespace \midrule
\multirow{3}{3cm}{\parbox{3cm}{Difficult data copyright protection, easy leakage and illegal copying}} & Data watermarking and fingerprinting & Embedding imperceptible watermarks in data & \cite{olaleye_composition_2022} \\ \cmidrule{3-4}
& & Identifying data sources through digital fingerprints & \cite{wu_privacyfriendly_2019}, \cite{sheng_cpchain_2020}, \cite{li_achieving_2022} \\ \midrule
\multirow{3}{3cm}{\parbox{3cm}{Data abuse and unauthorized use}} & Auditing and tracking mechanisms & Constructing data lifecycle auditing mechanisms, designing data element-oriented tracking technologies & \cite{yang_blockchainbased_2020}, \cite{zhang_privacypreserved_2024a} \\ \cmidrule{3-4}
& & Using blockchain to record data circulation trajectories & \cite{sheng_cpchain_2020}, \cite{olaleye_composition_2022} \\  \bottomrule
\end{tabular*}
\end{table*}

\begin{table*}[h!]
\centering
\caption{Data trading cost management methods}
\label{tab:cost_management}
\begin{tabular*}{\linewidth}{@{\extracolsep{\fill}}p{3cm}p{3cm}p{6cm}p{3cm}}
\toprule
Problem & Existing Strategies & Implementation Methods & Representative Literature \\ \midrule
\multirow{2}{2.5cm}{\parbox{2.5cm}{High trading costs due to data heterogeneity}} & Interoperability and standardization & Formulate unified standards for data formats, interfaces, etc. & \cite{elsaify_data_2020} \\ \cmidrule{3-4}
& & Develop semantic mapping, synchronization, and other cross-domain data interconnection technologies & \cite{koshizuka_dataex_2022}, \cite{iwasa_development_2020} \\ \midrule
\multirow{2}{2.5cm}{\parbox{2.5cm}{Difficulty in data pricing, lack of unified standards and rules}} & Data value assessment & Construct a multi-dimensional data pricing theoretical framework & \cite{shen_personal_2022}, \cite{liao_establishing_2023}, \cite{liang_data_2021} \\ \cmidrule{3-4}
& & Develop intelligent data pricing tools and platforms & \cite{hayashiTEEDAInteractivePlatform2020} \\ \bottomrule
\end{tabular*}
\end{table*}

\begin{longtable}{@{}p{2cm}p{2.5cm}p{3cm}p{7cm}p{2cm}@{}}
\caption{Summary of integrated data trading solutions}
\label{tab:method_summary} \\
\toprule
Problem & Existing Strategies & Key techniques & Implementation Methods & Representative Literature \\ \midrule
\endfirsthead
\caption{Summary of integrated data trading solutions (continued)} \\
\toprule
Problem & Existing Strategies & Key techniques & Implementation Methods & Representative Literature \\ \midrule
\endhead
\midrule \multicolumn{5}{r}{Continue to the next page} \\ \bottomrule
\endfoot
\bottomrule
\endlastfoot
\multirow{6}{2cm}{\parbox{2cm}{Compliance and Consequences}} & Construct trustworthy mechanisms & Blockchain,\newline smart contracts & Use blockchain to build traceability and evidence preservation mechanisms for credit records & \cite{calancea_techniques_2021} \\ \cmidrule{4-5}
& & & Use big data analysis to dynamically generate credit profiles & \cite{oh_deposit_2021} \\ \cmidrule{4-5}
& & & Use smart contracts to codify legal rules and realize automatic rule execution & \cite{luu_making_2016} \\ \cmidrule{2-5}
& Optimize data supply & Crowdsourcing,\newline machine learning,\newline social sensing & Build crowdsourcing task publishing platforms to support data collection, cleaning, labeling and other tasks & \cite{huang_efficient_2022},\cite{luu_making_2016},\cite{liang_survey_2018} \\ \cmidrule{4-5}
& & & Use big data analysis and machine learning to realize intelligent matching of tasks and participants and achieve data incentives & \cite{ramsundar_tokenized_2018a,gu_cbdtf_2024} \\ \midrule
\multirow{5}{2cm}{\parbox{2cm}{Compliance and Cost}} & Reshape trading architecture & Microservice architecture,\newline service orchestration & Develop data exchange platforms based on domain-driven design & \cite{hedenberg_developing_2018} \\ \cmidrule{4-5}
& & & Use containerization, service orchestration, and other technologies to improve the agility and trustworthiness of data services & \cite{niya_itrade_2021},\cite{sharma_trustworthy_2020} \\ \cmidrule{2-5}
& Construct decentralized payment systems & Decentralized currency,\newline cryptocurrency,\newline asymmetric encryption,\newline smart contracts,\newline consensus mechanisms & Construct decentralized currency systems based on blockchain, smart contracts and other technologies & \cite{lucking_when_2021},\cite{ramsundar_tokenized_2018a},\cite{oh_personal_2019} \\ \cmidrule{4-5}
& & & Map data assets to data currencies through smart contracts & \cite{xiong_smart_2019},\cite{chenli_fairtrade_2021} \\ \cmidrule{4-5}
& & & Use smart contracts, asymmetric encryption, consensus algorithms, etc., to ensure currency circulation & \cite{xiong_data_2021} \\ \cmidrule{2-5}
& Construct consensus mechanisms & Consensus algorithms,\newline incentive mechanisms & Use blockchain to build a distributed governance architecture and form a pattern of multi-stakeholder participation & \cite{tian_optimal_2019} \\ \midrule
\multirow{4}{1.5cm}{\parbox{1.5cm}{Cost and Consequences}} & Improve incentive mechanisms & Federated learning,\newline incentive mechanisms & Design federated learning architectures to form multi-party participation & \cite{hu_trading_2020},\cite{zheng_flmarket_2022} \\ \cmidrule{4-5}
& & & Design incentive mechanisms to mobilize the enthusiasm of all parties to participate & \cite{li_novel_2022},\cite{khan_federated_2020} \\ \cmidrule{2-5}
& Optimize data management & Metadata management,\newline cloud computing & Design a unified metadata framework & \cite{lawrenz_significant_2020},\cite{hayashi_understanding_2020} \\ \cmidrule{4-5}
& & & Build data management platforms to support full lifecycle data management & \cite{zhou_securing_2023} \\ \bottomrule
\end{longtable}

\end{document}